\documentclass[aps,prd,showpacs,amssymb,amsmath,amsfonts,
twocolumn,floatfix,preprintnumbers]{revtex4}
\usepackage{graphicx}
\usepackage[raggedright,hang]{subfigure}
\usepackage{epstopdf}
\usepackage{rotating}
\usepackage{color}
\usepackage{hyperref}
\usepackage{amssymb}
\usepackage{amsmath}
\usepackage{fancyhdr}

\usepackage{dcolumn}
\usepackage{bm}

\begin{document}
\pagestyle{fancy}
\rhead[]{}
\lhead[]{}

\title{Gravitational waves from the $r$-mode instability of neutron stars: effect of magnetic field }

\author{Bhim Prasad Sarmah}
\email{bhim@tezu.ernet.in}
\affiliation{%
Department of Mathematical Sciences, Tezpur University, Napaam-784001, India} 
\author{H L Duorah}
\affiliation{%
Department of Physics, Gauhati University, Gopinath Bardoloi Nagar-781014, India}

\begin{abstract}
Studies have shown that emission of gravitational waves drives an instability in the $r$-modes of young rapidly rotating neutron stars carrying away most of the angular momentum through gravitational wave emission in the first year or so after their formation. Magnetic field plays a crucial role in the evolution of these $r$-modes and hence the evolution of the neutron star itself. An attempt is made here to investigate the role of magnetic field in the evolution of $r$-mode instability and detectibility of gravitational waves emitted by a newly born, hot and rapidly and differentially rotating neutron star. It is found that magnetic fields tend to suppress the $r$-mode amplitude. The {\it signal-to-noise ratio} analysis shows that gravitational waves emitted from the $r$-mode instability from neutron stars with magnetic fields upto the order of $10^{14}$ gauss may be detectable by the Advanced LIGO at 20 Mpc.
\end{abstract}


\maketitle

\section{Introduction}

Recent developments in the kilometer$-$scale laser interferometric gravitational$-$wave detectors LIGO and Virgo, or their advanced versions, operating in the frequency range of 10 and 10$^4$ Hz usher a new era in the gravitational-wave astronomy, enhancing the prospects of detecting  gravitational waves from a variety of astrophysical sources.  

The $r$-modes are primarily non-radial velocity perturbations in stars. These perturbations are of the order of the angular velocity $\Omega$ of stars' rotation and entail density perturbations of order $\Omega^2$. The characteristic frequency of oscillation is also comparable to $\Omega$. 

The interesting fact about the $r$-modes is that they are driven unstable by gravitational radiation reaction in all perfect-fluid rotating stars for a large range of relevant angular velocities. This property was discovered by Andersson\cite{andersson98} and confirmed theoretically by Friedman and Morsink\cite{fried_morsnk98}. So, unlike other modes, $r$-modes are always retrograde in the star's corotating frame and prograde in the inertial frame, i.e., the sign of the $r$-mode frequencies is always opposite in the two frames. In other words the Chandrasekhar-Friedman-Schutz (CFS) instability always occurs for the $r$-modes.

 The effects of bulk and shear viscosity in neutron star matters are to dampen the driving effects of gravitational radiation. It is found that in case of neutron star matters, the effect of gravitational radiation driven instability is stronger than the damping effects of viscosity \cite{lind_owen_morsnk98,and_kok_schutz99}.

A phenomenological model for evolution of the $r$-mode instability was given by Owen {\it et. al.}\cite{OLCSVA} to study the detectibility of gravitational waves emitted by a newly born, hot, rapidly-rotating neutron star. They found that such waves could be detected by the enhanced version of LIGO if sources were located at distances up to 20 Mpc from Earth. However, a deeper understanding of this issue requires taking into account of nonlinear effects  in the evolution of $r$-mode instability. Arras {\it et al.} \cite{arras} studied the nonlinear mode interaction to conclude that enhanced laser interferometric detectors could detect gravitational radiation from $r$-modes provided the sources are located within a distance of 200kpc. That $r$-modes can induce a drift of fluid elements along azimuthal directions was suggested by Rezzolla {\it et al.}\cite{rezzlett,rezz1} who derived an approximate analytical expression for these drifts and these were numerically confirmed in both general relativistic \cite{stergfont} and Newtonian Hydrodynamics\cite{lindtohval1,lindtohval2} regimes.

Recently, S\'a \cite{sa1} has found an exact $r$-mode solution within a nonlinear theory up to second order in the mode's amplitude which describes differential rotation of pure kinematic nature producing large scale drifts along stellar latitudes. This solution is then used to investigate the influence of differential rotation in the evolution of $r$-mode instability of a newly born, hot, rapidly rotating neutron star \cite{sa2} and the detectibility of gravitational waves emitted by such a system by Advanced laser interometric detector LIGO \cite{sa6}.

A newly born hot neutron star is beset with high spin and coherent, large-scale hot plasma fluid currents with high electrical conductivity.  In such neutron star matter, large scale magnetic fields can be produced. It will thus be interesting to investigate the influence of magnetic field in the evolution of newly born hot neutron star, in particular its influence on the  spin evolution through $r$-mode instability. Spruit \cite{spruit99} first investigated the generation of a toroidal magnetic field as a result of gravitational wave emission induced differential motion in the star. The huge magnetic field produced along with the buoyancy instability associated with them have been used for the production of a gamma-ray burst model. Ho and Lai \cite{holai2000} introduced magnetic braking originated from the magnetic dipole radiation in the spin evolution of a unstable neutron star and estimated  the growth rate of the instability as a result of fast magnetosonic and Alfv\'en wave emission in the star's magnetosphere. 

The present work attempts to make phenomenological descirption of the nonlinear evolution of the $r$-modes in a newly born hot neutron star up to second order in mode's amplitude by extending the work of S\'a and investigate the influence of the magnetic field on the mode amplitude and the kinematic differential rotation. 


In section II we describe briefly the general properties of the $r$-mode, section III details evolution of $r$-modes in a canonical neutron star with or without invoking the differential rotation and magnetic fields. In section IV we calculate the waveform of gravitational waves emitted during $r$-mode evolution. The results are analysed in section V. We summarise our results in section VI.  

\section{The r-mode instability: General consideration}
The Eulerian velocity perturbation in the rotating Newtonian stars in case of r-modes are given by 
\begin{equation}
\delta\vec{v}=R~\Omega~f(r/R)~\vec{Y}_{lm}^B~e^{ iwt } 
\end{equation}
where $R$ and $\Omega$ are the radius and angular velocity of the unperturbed star, $f(r/R)$ is an arbitrary dimensionless function, and $\vec{Y}_{lm}^B$ is the magnetic type vector spherical harmonic defined by
\begin{equation}
\vec{Y}_{lm}^B=[l(l+1)]^{-1/2}r\vec\nabla\times(r\vec\nabla {Y}_{lm})
\end{equation}

For barotropic models, the radial dependence $f(r/R)$ equals $\alpha (r/R)^l$, where $\alpha$ is an arbitrary constant \cite{ProvBerthRocca}. These modes exist with velocity perturbations as given by Eq. (1) if $l=m$, and the mode frequencies are 
\begin{equation}
\omega=-\frac{(l-1)(l+2)}{l+1}\Omega
\end{equation}
Since the Coriolis force serves the restoring force for these oscillations, the frequencies of these modes are low compared to the $f$ and $p$ -modes in slowly rotating stars. These expressions for $ \delta\vec{v}$ and ${\omega}$ are the lowest order in an expansion in terms of the angular velocity $\Omega$. The exact expressions contain additional terms of the order $\Omega^3$. 


The form of the $r$-mode solution has the time dependence $e^{iwt-t/\tau}$ as a consequence of ordinary hydrodynamics and the influence of the various dissipative processes. The real part is the frequency of these modes, $\omega$, given in $(3)$, while the imainary part $1/\tau$ is determined by the effects of gravitational radiation, viscosity, etc. It is possible to evaluate $1/\tau$ by computing the time derivative of the energy $\overline E$ in the rotating frame and the time derivative of $\overline E$. The energy $\overline E$ can be expressed as a quadratic functional of fluid velocity perturbations as:
\begin{equation}
\overline E=\frac{1}{2}\int \left[\rho \delta\vec v.\delta\vec v^*
+\left(\frac{\delta p}{\rho}-\delta\Phi\right)\delta\rho^*\right]d^3x
\end{equation}  
and
\begin{equation}\label{dedt}
\frac{d\overline E}{dt}=-\frac{2\overline E}{\tau}.
\end{equation}

We can decompose $1/\tau$ as
\begin{equation}
\frac{1}{\tau(\Omega)}=\frac{1}{\tau_{GR}(\Omega)}+\frac{1}{\tau_S(\Omega)}+\frac{1}{\tau_B(\Omega)},
\end{equation}
where $1/\tau_{GR}, 1/\tau_S$ and $1/\tau_B$ are the contributions due to gravitational radiation emission, shear viscosity and bulk viscosity, respectively. Expressions for these individual contributions for  the $r$-modes are given by \cite{lind_owen_morsnk98}
\begin{equation}\label{tauGR}
\frac{1}{\tau_{GR}}=-\frac{32\pi G \Omega^{2l+2}}{c^{2l+3}}
\frac{(l-1)^{2l}}{\left[\left(2l+1\right)!!\right]^2}
\left(\frac{l+2}{l+1}\right)^{2l+2}\int_{0}^{R}\rho r^{2l+2} dr,\\
\end{equation}
\begin{equation}\label{tauS}
\frac{1}{\tau_S}=(l-1)(2l+1)\int_0^R \eta r^{2l}dr \left(\int_0^R \rho  r^{2l+2}dr\right)^{-1}  
\end{equation}
and
\begin{equation}\label{tauB}
\frac{1}{\tau_B}\approx\frac{4R^{2l-2}}{(l+1)^2}\int \zeta\left|\frac{\delta\rho}{\rho}\right|^2d^3x\left(\int_0^R\rho r^{2l+2}dr\right)^{-1} ,
\end{equation}
where $\delta\rho$ is proportional to $\Omega^2$ and hence is small (i.e., higher order in $\Omega$) compared to $\delta\vec v$ in slowly rotating stars. 

The expression for $1/\tau_B$ in Equation (\ref{tauB}) is only approximate. The exact expression should contain the Lagrangian density perturbation $\Delta\rho$ in place of the Eulerian perturbation $\delta\rho$. The bulk viscosity  [Eq.(\ref{exzeta})] is a very strong function of the temperature, being proportional to $T^6$.  

Owen et. al.(1998) \cite{OLCSVA} have evaluated these expressions  for the imaginary parts of the frequency for a ``typical" neutron star model with a polytropic equation of state : $p=k\rho^2$, with $k$ chosen so that a $1.4M_\odot$ model has the radius $12.53$ km by using the expressions for the viscosity of the hot neutron star matter \cite{sawyer89, cutlind87} :
\begin{equation}\
\eta=347\rho^{9/4}T^{-2}
\end{equation}
\begin{equation}\label{exzeta}
\zeta=6.0\times10^{-59}\left(\frac{l+1}{2\Omega}\right)^2\rho^2T^6
\end{equation}
where all quantities are expressed in cgs units. 

It will be useful to define a timescale associated with the viscous dissipation $1/\tau_V=1/\tau_S+1/\tau_B$. The viscous timescale $\tau_V$ and the gravitational timescale $\tau_{GR}$ can be expressed in terms of the temperature and angular velocity dependences as 
\begin{equation}\label{tauV}
\frac{1}{\tau_V}=\frac{1}{\widetilde{\tau}_S}\left(\frac{1}{T_9}\right)^2 
+\frac{1}{\widetilde{\tau}_B} T_9^2 \left( \frac{\Omega^2}{\pi G \overline{\rho}}\right)
\end{equation}
\begin{equation}\label{tauGRfnl}
\frac{1}{\tau_{GR}}=\frac{1}{\widetilde{\tau}_{GR}}\left(\frac{\Omega^2}{\pi G \overline{\rho}}\right)^{l+1}
\end{equation}
where $\widetilde{\tau}_S=2.52\times10^8$ seconds, $\widetilde{\tau}_B=6.99\times10^8$ seconds and $\widetilde{\tau}_{GR}=-3.26$ seconds for $l=2$ mode \cite{OLCSVA}.
\section {Evolution of the r-modes in neutron stars}
\subsection {In the absence of differential rotation and magnetic field}

In the consideration of the evolution of the $r$-modes in neutron star environments, it is assumed that initially the mode will be a small perturbation that is described adequately by the linear analysis. However, as the mode grows, non-linear hydrodynamic effects become important and eventualy dominate the dynamics. In the absence of the mathematical tool to exactly follow this non-linear phase of the evolution, Owen et. al. \cite{OLCSVA} first proposed a simple linearised treatment that provides the basic features of the exact evolution. Essentially the neutron star is treated as a simple system having only two degrees of freedom: the uniformly rotating equilibrium state parametrized by its angular velocity $\Omega$, and the $r$-mode is parametrized by its amplitude $\alpha$. 

The canonical angular momentum, {\it i.e.}, the angular momentum associated with the $r$-mode,  
can be expressed in terms of the velocity perturbation $\delta\vec{v}$ by 
\begin{equation}
J_c=-\frac{l}{2(\omega+l\Omega)}\int \rho\delta\vec{v}.\delta\vec{v}^*d^3x
\end{equation}
For the $l=2$ $r$-mode, this expression at the lowest order in $\Omega$ reduces to 
\begin{equation}\label{jc}
J_c=-\frac{3}{2}\Omega\alpha^2\widetilde{J}MR^2
\end{equation}
where $\widetilde{J}$ is given by 
\begin{equation}\label{jtilde}
\widetilde{J}=\frac{1}{MR^4}\int_0^R \rho r^6 dr
\end{equation}
For polytropic models the dimensionless constant $\widetilde{J}=1.635\times10^{-2}$. The moment of inertia $I$ can be  written as 
\begin{equation}\label{mom_inertia}
I=\widetilde{I}MR^2
\end{equation}
where $\widetilde{I}$ is given by
\begin{equation}\label{itilde}
\widetilde{I}=\frac{8\pi}{3MR^2} \int_0^R \rho r^4 dr
\end{equation}
and for $\Gamma=2$ polytropic models,  $\widetilde{I}=0.261$. 

The angular momentum associated  with the mean rigid rotation primarily increases  through a transfer from the mode due to viscosity through the emission of gravitational radiation.  Thus,
\begin{equation}\label{iomgrate}
\frac{d(I\Omega)}{dt}=\frac{2}{\tau_V}J_c
\end{equation}

The rate of change of canonical momentum of the mode can increase through gravitational radiation and decrease by transferring angular momentum to the star through viscosity. Consideration of this fact leads to another evolution equation for the $r$-modes. This was realised by Ho and Lai \cite{holai2000} that it is the evolution of the canonical momentum and not the energy evolution as in Owen et. al.\cite{OLCSVA},  that is responsible for the correct description of the evolution of r-mode. 
Thus, 
\begin{equation}\label{jcrate} 
\frac{dJ_c}{dt}=-\frac{2}{\tau_{GR}}J_c-\frac{2}{\tau_V}J_c
\end{equation}
Substituting equations (\ref{jc}) through (\ref{itilde}) into equation (\ref{iomgrate}) yields 
\begin{equation}
\frac{d\Omega}{dt}=-2Q\frac{\Omega \alpha^2}{\tau_V}
\end{equation}
where $Q=3\widetilde{J}/2\widetilde{I}=9.40\times 10^{-2}$. 
\begin{equation}
 \frac{d\alpha}{dt}=-\frac{\alpha}{\tau_{GR}}-\frac{\alpha}{\tau_V}(1-\alpha^2 Q)
\end{equation}
for the early stage,  and
\begin{center}
$\alpha^2=\kappa$\\
\end{center}
and \
\begin{equation}
\frac{d\Omega}{dt}=\frac{2\Omega}{\tau_{GR}}\frac{\kappa Q}{1-\kappa Q}
\end{equation}
for later stage (after saturation). 

\subsection {$r$-mode evolution of differentially rotating neutron stars but no magnetic field}
Sa et. al. \cite{sa1,sa2} while attempting to find out the nonlinear correction (in the second order in the mode's amplitude $\alpha$), to the equations of Owen et. al, found that the leading non-linear effect can be interpreted as differential rotation only if one includes a correction term to the expression of the canonical angular momentum. After a detailed nonlinear calculation, they found that the corrected version of the canonical angular momentum term (called as physical angular momentum of the $r$-mode) is given by \cite{sa2}

\begin{equation}
\delta^{(2)} J =\frac{1}{2}\alpha^2\Omega(4K+5)\widetilde{J}MR^2
\end{equation}
 where $\Omega, R$  and $M$ denote the angular velocity, the radius and the mass of the star, respectively so that the total angular momentum of $r$-mode is 
\begin{equation}
J=I\Omega+\delta^{(2)} J
\end{equation}
and $K$ is a parameter accounting for the differential rotation in the $r$-mode evolution. 

Incorporation of the above correction leads to the $r$-mode evolution equation for a differentially rotating star \cite{sa2}
\begin{equation}
\frac{d\Omega}{dt}=\frac{8}{3}(K+2)Q\alpha^2\frac{\Omega}{\tau_{GR}}+\frac{8}{3}\left(K+\frac{5}{4}\right)Q\alpha^2\frac{\Omega}{\tau_V}
\end{equation}
\begin{equation}
-\left[1+\frac{4}{3}(K+2)Q\alpha^2\right]\frac{\alpha}{\tau_{GR}}                                     
 -\left[1+\frac{4}{3}\left(K+\frac{5}{4}\right)Q\alpha^2\right]\frac{\alpha}{\tau_V}
\end{equation}
It is to be noted that the evolution equations for a rigidly rotating star (similar to \cite{OLCSVA} and also \cite{holai2000} with the suppression of the magnetic field term) can be recovered from the above equations by setting $K=-2$.  
Another advantage with this set of evolution equation is that they account for the entire history of $r$-mode evolution - one need not monitor the saturation value of $\alpha$ and then change the governing equation as had to be done in the case of either Owen et. al or of Ho and Lai.   
\subsection{$r$-mode evolution of differentially rotating neutron star in the presence of magnetic field}

As the neutron star is beset with a very strong magnetic field,  of the order of $10^{14-16}$ Gauss , the role of such a huge magnetic field cannot be ruled out in governing the evolution of r-mode. This is quite possible that evaluation of different other quantity like the amplitude of gravitational wave that is emitted by such a neutron star during $r$-mode evolution period and the amount of spin down and spin down period etc. will be overestimated if the magnetic field is not accounted for during  $r$-mode evolution of a neutron star. Ho and Lai incorporated the effect of magnetic field in a very simple way by considering the magnetic braking effect during the spin down of the star. Though a detail analysis of the role of magnetic field is itself a formidable task, in this paper we incorporate the magnetic field into the evolution equation of Sa et. al. in the same line as done by Ho and Lai and analyse its effect in the differential rotation of the star during $r$-mode evolution. We further calculate the amount of gravitational radiation to be emitted by a neutron star in the newtonian regime with a polytropic equation of state and for different strengths of magnetic fields and differential rotation parameter to have a qualitative estimate of the role played by the magnetic field and differential rotation simultaneously in the gravitational wave emission. We further look into the gravitational wave detectibility aspect of such a star in the eyes of LIGO I, LIGO II and VIRGO.\\

In the  $l=2\;\;r$-mode instability of second order non-linear theory, two two crucial parameters are the angular velocity $\Omega$ and the mode amplitude $\alpha$. The physical angular momentum of the second order perturbation, which includes the differential rotation in a typical neutron star is given by 

\begin{equation}\label{ddel2jdt}
\delta^{(2)}J=\frac{1}{2}\alpha^2\Omega(4K+5)\widetilde{J}MR^2
\end{equation}

This expression assumes the star's mass density $\rho$ and pressure $p$ to be related by a polytropic equation of state $p=k\rho^2$ and the constant $k$ is chosen such that $M=1.4M_\odot$ and $R=12.53$ kilometres. The constant $K$ in equation (\ref{ddel2jdt}) is to be fixed by initial data, giving the initial amount of differential rotation associated with the $r$-mode \cite{sa1}.

The total angular momentum is the sum of the bulk angular momentum and the mode angular momentum, or the angular momentum of oscillation, is given by

\begin{equation}\label{jtot}
J=I\Omega+\delta^{(2)}J=\frac{1}{2}\alpha^2\Omega(4K+5)\widetilde{J}MR^2
\end{equation}

\vspace{.25cm}
where $I=(8\pi/3)\int_{0}^{R} \rho r^2 dr=\widetilde{I}MR^2$, ($\widetilde{I}=0.261$) is the moment of inertia of the unperturbed star. 

The angular momentum of oscillation $\delta^{(2)}J$ increases mainly due to the emission of gravitational wave and the dissipative effects of bulk viscosity and shear viscosity. Taking into consideration of these facts we find 

    \begin{equation}
    \frac{d\delta^{(2)}J}{dt}=-\frac{2}{\tau_{GR}}\delta^{(2)}J-\frac{2}{\tau_V}\delta^{(2)}J
    \end{equation}

On the other hand, the total angular momentum $J(\alpha,\Omega)$ decreases due to emission of gravitational wave and Magnetic braking in the case of a star beset with magnetic field of strength B. This gives
     \begin{equation}
       \frac{dJ}{dt}=3 \widetilde{J}MR^2 \frac{\alpha^2\Omega}{\tau_{GR}}-\frac{\widetilde{I}MR^2\Omega}{\tau_M}
      \end{equation}

The magnetic braking term arises from the radiation of rotational energy from the rotating magnetic neutron star over a time scale $\tau_M$ due to the intervention of magnetic field and leads to an equivalent magnetic braking torque of magnitude $-I\Omega/\tau_M$. For a simple magnetic dipole model of neutron star, we have \cite{shap-teuk}

\begin{equation}\label{tauM}
\frac{1}{\tau_M}=\frac{B^2R^6\Omega^2}{6c^3I}
\end{equation}

Thus the consideration of decrease of total angular momentum due to emission of gravitational wave and magnetic braking, leads to the equation

\begin{equation}\label{djdt}
\frac{dJ}{dt}=3\widetilde{J}MR^2\frac{\alpha^2\Omega}{\tau_{GR}}-\frac{\widetilde{I}MR^2\Omega}{\tau_M}
\end{equation}

Using equation (\ref{jtot}), equation (\ref{djdt}) reduces to 
\begin{widetext}
\begin{equation}\label{domgdt}
\frac{d\Omega}{dt}\left[1+\frac{1}{3}Q(4K+5)\alpha^2\right]+\frac{2}{3}Q(4K+5)\alpha\Omega\frac{d\alpha}{dt}=\frac{2Q\alpha^2\Omega}{\tau{GR}}-\frac{\Omega}{\tau_M}
\end{equation}
\end{widetext}

where 
\begin{equation}
Q=\frac{3\widetilde{J}}{2\widetilde{I}}\nonumber
\end{equation}

On the other hand, consideration of the increase of $\delta^{(2)}J(\alpha, \Omega)$ due to gravitational radiation instability and decrease due to bulk and shear viscosity, leads to 

\begin{widetext}
\begin{equation}
\frac{d}{dt}\left[\frac{1}{2}\alpha^2\Omega(4K+5)\widetilde{J}MR^2\right]=-\alpha^2\Omega(4K+5)\widetilde{J}MR^2\left(\frac{1}{\tau_{GR}}+\frac{1}{\tau_V}\right)
\end{equation}

On simplification, 

\begin{equation}\label{daldtfnl}
\frac{d\alpha}{dt}=-\frac{\alpha}{\tau_{GR}}\left[1+\frac{4}{3}Q\alpha^2(K+2)\right]-\frac{\alpha}{\tau_V}\left[1+\frac{\alpha^2Q(4K+5)}{3}\right]+\frac{\alpha}{2\tau_M}
\end{equation}
\end{widetext}

which, when used in equation (\ref{domgdt}), gives

\begin{equation}\label{domgdtfnl}
\frac{d\Omega}{dt}=\frac{8\alpha^2\Omega Q (K+2)}{3\tau_{GR}}+\frac{2(4K+5)\alpha^2\Omega Q}{3\tau_V}-\frac{\Omega}{\tau_M}
\end{equation}

Equations (\ref{daldtfnl}) and (\ref{domgdtfnl}) are the governing equations for evolution of $r$-mode instability in the presence of both differential rotation and magnetic field. $K$ is a constant which is to be fixed by the choice of initial data, giving the initial amount of differential rotation associated with the $r$-mode. The value of $K$ is chosen to lie in the interval $-5/4 \leq K \ll 10^{17}$. The upper limit for $K$ results from the imposition of the condition that the initial value of the angular momentum of the $r$-mode is much smaller than the angular momentum of the unperturbed star, {\it i.e.}, $\delta^{(2)}\ll I\Omega_0$, which gives $K \ll 10^{17}$, if we choose $\alpha_0=10^{-8}$. The lower limit for $K$ results from the fact that we want to avoid the total angular momentum of the star becoming a negative quantity. This will further ensure that amplitude saturation is governed by the equations themselves unlike the case in \cite{OLCSVA} where the saturation of the $r$-mode amplitude was put by hand.

The equations(\ref{daldtfnl}) and (\ref{domgdtfnl}) are numerically solved 
for a range of values of initial differential rotation and magnetic field strengths. Convergence of the numerical solutions are checked by changing the values of the step-size of iterations and found to be numerically convergent. For the purpose of illustrating the behaviour of nonlinear growth, we have chosen the values of differential rotation parameter $K$ and the magnetic field strengths as the following:

\begin{table}[h]
\begin{tabular}{c|ccccccccccccc}
\hline\hline
&&&&&&&&&&&&& \\
$K$ & 0 & 10 & 100 & 1000 & $10^4$ & $10^5$ & $10^6$ & $10^7$ & $10^8$ & $10^9$ & $10^{10}$ & $10^{11}$ & $10^{12}$ \\[6pt]
\hline
&&&&&&&&&&&&& \\
$B_{14} $ & 0 & 2 & 4 & 6 & 8 & 10 & 15 & 20 & 40 & 60 & 80 & 100 & \\[6pt]
\hline\hline
\end{tabular}
\caption{Table of different values of initial differential rotation parameter $K$ and the magnetic field $B_{14}$ ($\times 10^{14}$ Gauss) chosen for the numerical solution of Equations (\ref{daldtfnl}) and (\ref{domgdtfnl}).}\label{b_k_table}
\end{table}

$B_{14}$ is the value of the magnetic field strength in units of $10^{14}$ Gauss. For different combinations of the values of $K$ and $B_{14}$ chosen above, we have numerically solved the equations and analysed the results for the growth of mode amplitude, evolution of angular velocity, gravitational wave amplitude and the detectibility of emitted gravitational wave signals. 

It is further necessary to specify how the temperature of the star evolves with time in order to make the model of the evolutionnof the $r$-modes complete. We adopt here the standard description of the cooling of hot young neutron stars due to emission of neutrino via the modified URCA process. The temperature during this phase falls quickly by a simple power law cooling formula\cite{shap-teuk},

\begin{equation}\label{temp_time}
\frac{T(t)}{10^9 K}=\left[\frac{t}{\tau_c}+\left(\frac{10^9 K}{T_i}\right)^6\right]^{-\frac{1}{6}}
\end{equation}

where $T_i$ is the initial temperature of the neutron star, and $\tau_c$ is a parameter that characterises the cooling rate. For the modified URCA process, $\tau_\approx$ 1 year. A typical value for the initail temperature is $T_i\approx 10^{11}$ K. This equation can now be inserted into the evolution equations for $\alpha$ and $\Omega$ to provide explicit differential equations fo rthe time evolution of the angular velocity of the star and the amplitude of the mode. 

\begin{figure}[htp]
\includegraphics[width=0.5\textwidth, height=0.3\textheight]{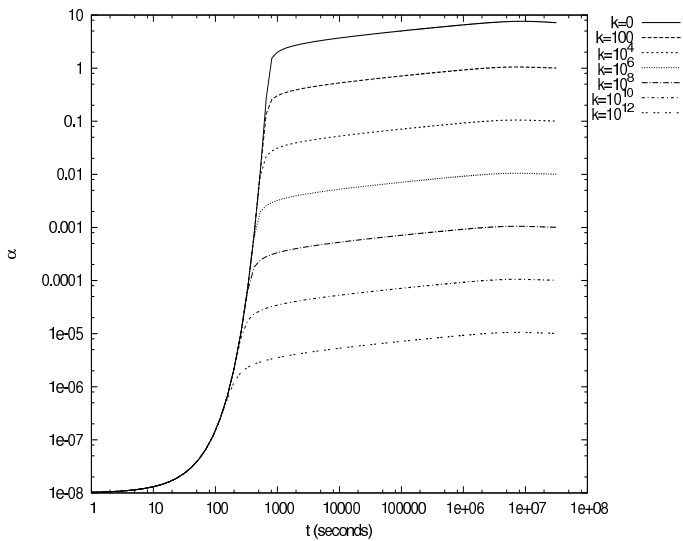}
\caption{Growth of r-mode amplitude for different values of the differential rotation parameter K, and magnetic field B=0.}\label{alpha_time_b0_k.lbl}
\end{figure}  


\begin{figure}[htp]
\includegraphics[width=0.5\textwidth, height=0.3\textheight]{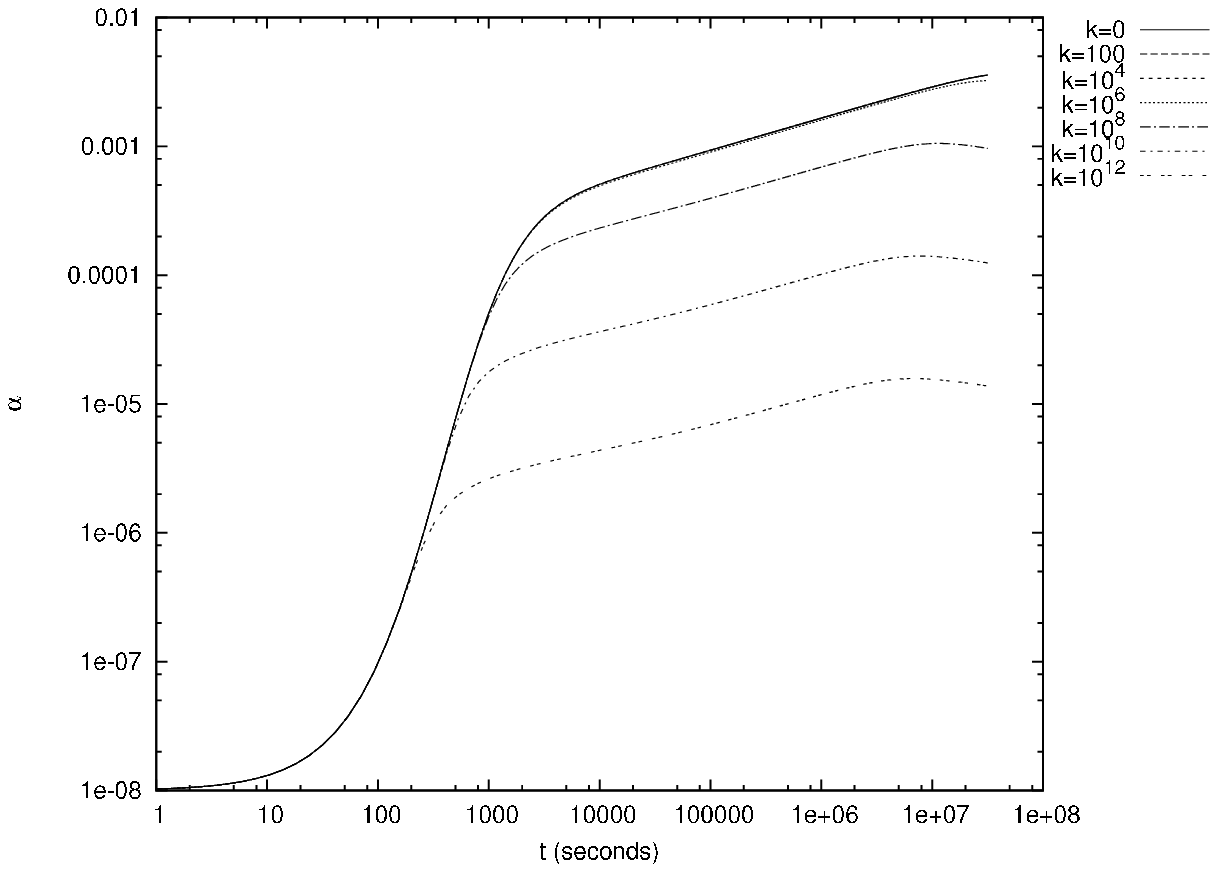}
\caption{Growth of r-mode amplitude for different values of the differential rotation parameter K, and magnetic field $B_{14}=10$. }
\end{figure}  

\begin{figure}[htp]
\includegraphics[width=0.5\textwidth,height=0.3\textheight]{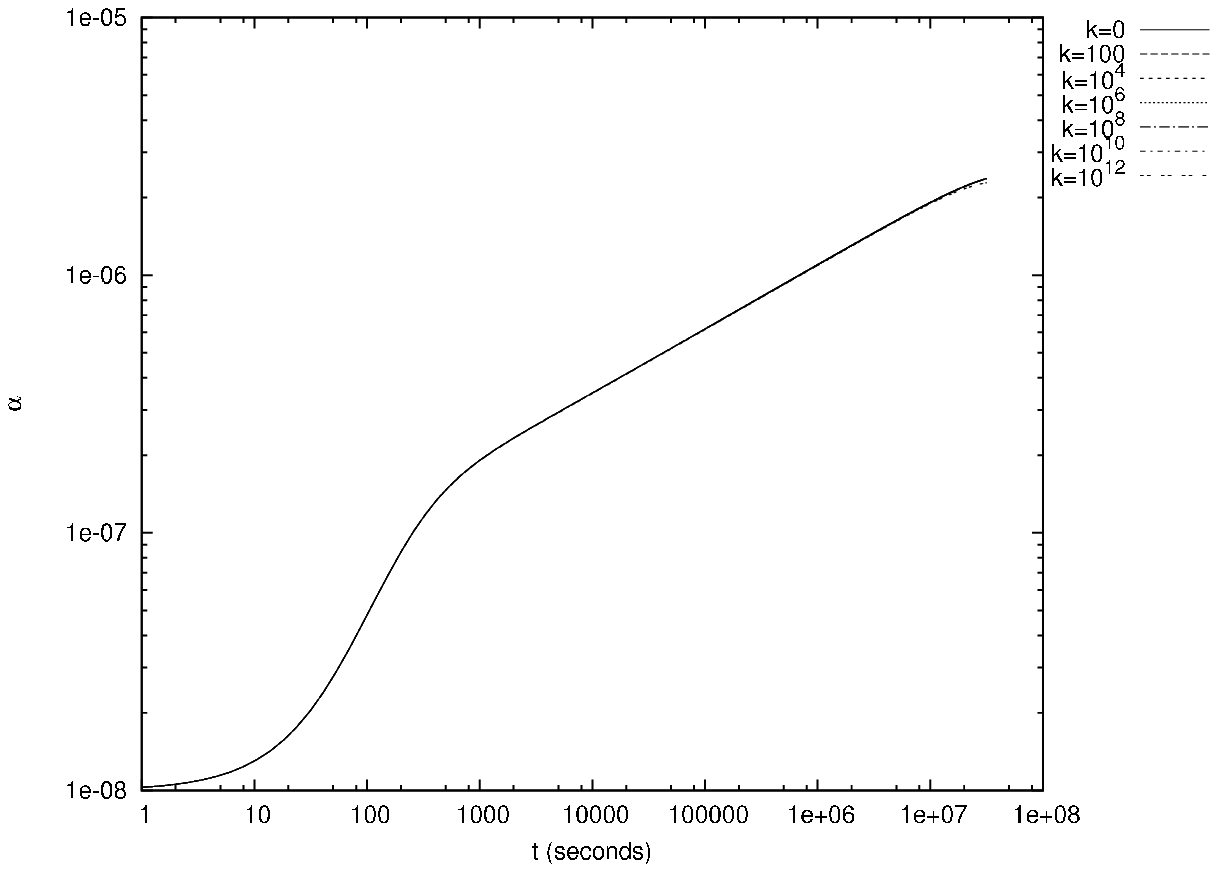}
\caption{Growth of r-mode amplitude with time for different values of K, the differential rotation parameter and with magnetic field $B_{14}=20$. }\label{alpha_time_b20_k.lbl}
\end{figure}  

\begin{figure}[htp]
\includegraphics[width=0.5\textwidth,height=0.3\textheight]{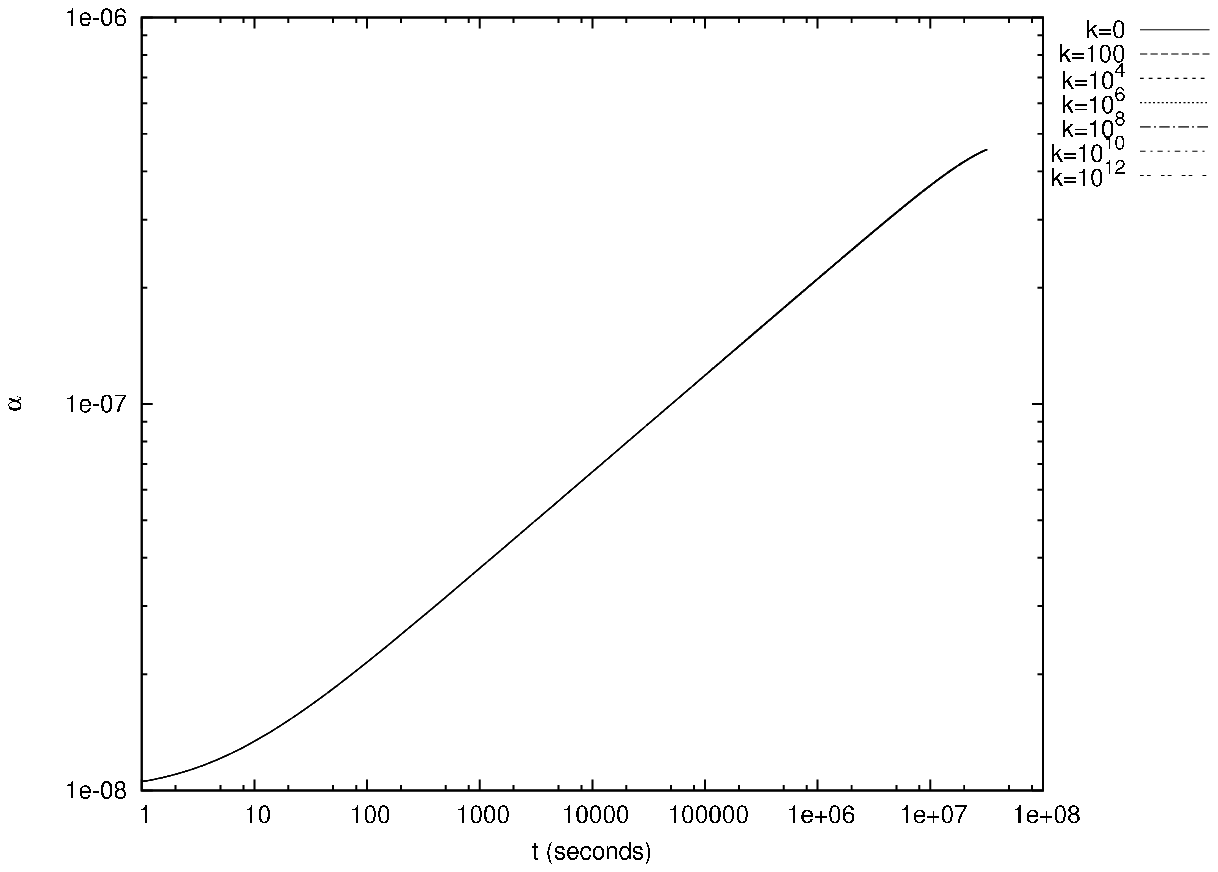}
\caption{Growth of r-mode amplitude with time for different values of K, the differential rotation parameter and with magnetic field $B_{14}=100$. }
\end{figure}

\begin{figure}[htp]
\includegraphics[width=0.5\textwidth,height=0.3\textheight]{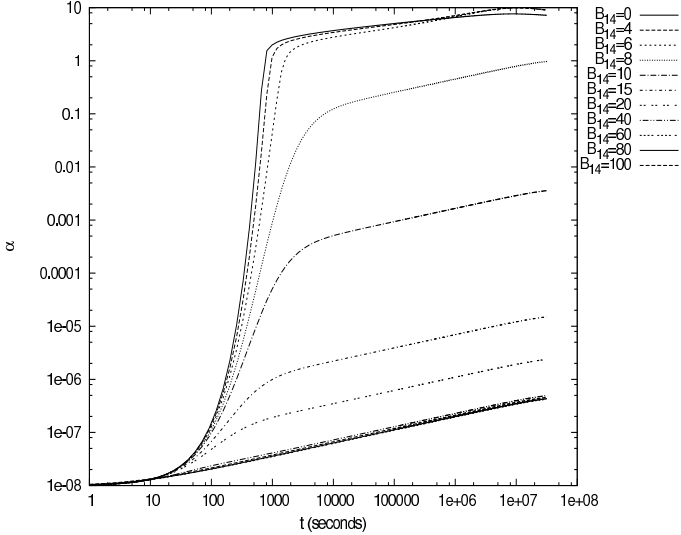}
\caption{Growth of r-mode amplitude with time for different values of magnetic field and with differential rotation parameter K=0. }
\end{figure}  


\begin{figure}[htp]
\includegraphics[width=0.5\textwidth,height=0.3\textheight]{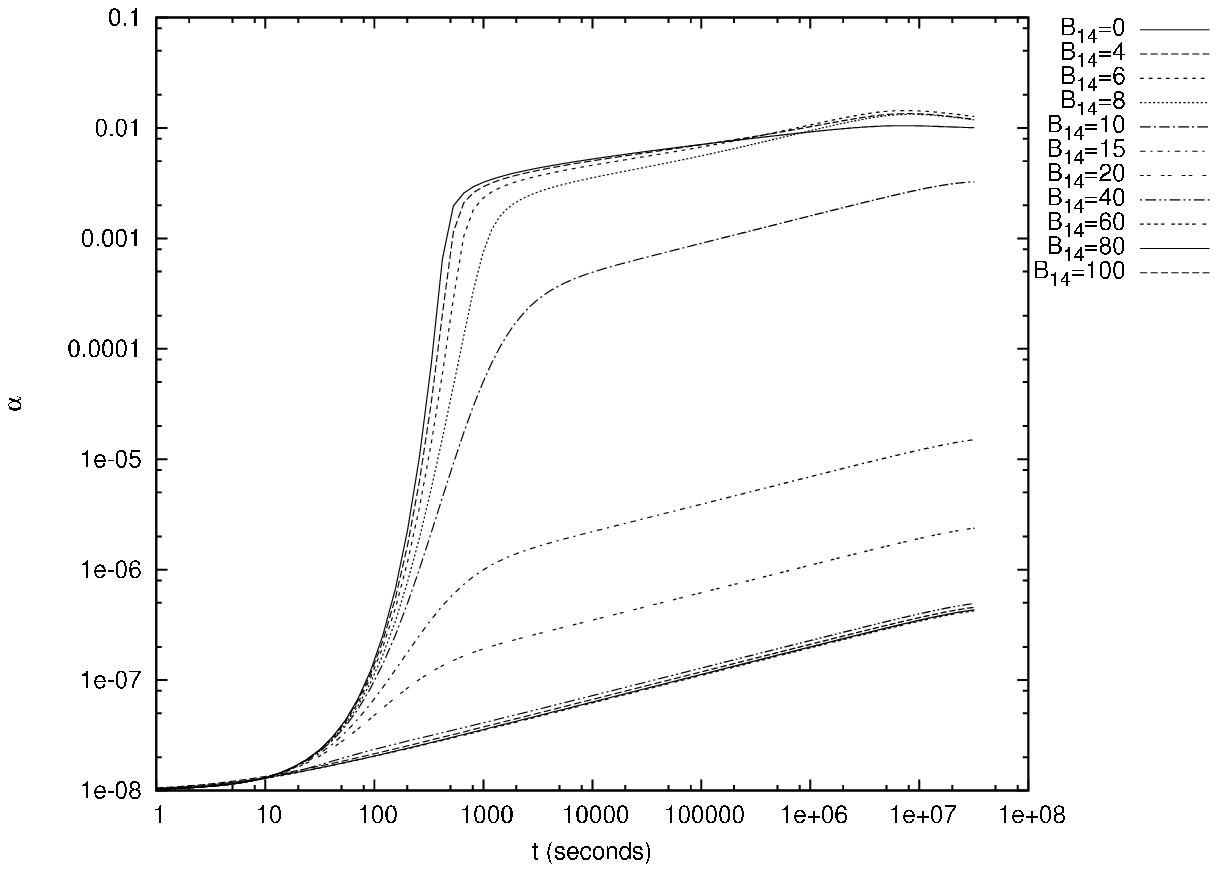}
\caption{Growth of r-mode amplitude for different B and $K=10^6$.}
\end{figure}


\begin{figure}[htp]
\includegraphics[width=0.5\textwidth,height=0.3\textheight]{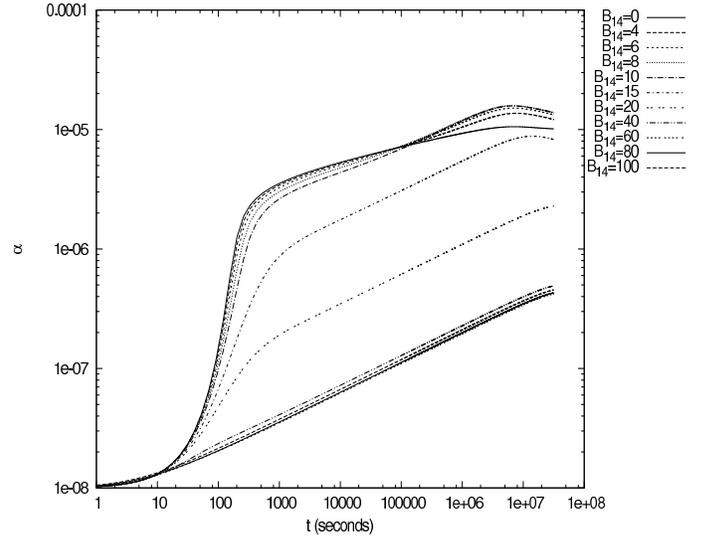}
\caption{Growth of r-mode amplitude for different B and $K=10^{12}$.}
\end{figure}           

\section{Gravitational waveform from r-mode instability in the presence of differential rotation and magnetic field and their detectibility}
The frequency-domain gravitational waveform
\begin{equation}
\widetilde{h}(f)\equiv \int_{-\infty}^{\infty} e^{2\pi i f t}h(t)dt
\end{equation}
is determined completely by the assumption that the angular meomentum radiated as gravitational waves comes directly from (the angular momentum of) the star. This assumption is expected to be satisfied during the non-linear saturated phase of the evolution, but not during the early evolution when the mode is growing exponentially. This follows that the rate of the frequency of the radiation evolves with time in the later phase of the evolution.

In the stationary phase approximation, the gravitational wave strain $h(t)$ is related to its Fourier transform $\widetilde{h}(f)$ by
\begin{equation}
|h(t)|^2=|\widetilde{h}(f)|^2 \; \Big|\frac{df}{dt}\Big|
\end{equation}

For the $l=2$ of the $r$-mode which we discuss here, the mode frequency is $\omega=\frac{4}{3}\Omega$, or $f=\frac{2}{3\pi}\Omega$, where $f$ is the frequency of the emitted gravitational waves measured in Hz. 

The rate of angular momentum loss due to the $r$-mode can be coupled to the emission of gravitational waves in the standard current multipole expansion for $l=m=2$ as
\begin{equation}
\frac{dJ}{dt}=-\frac{c^3}{16\pi G}\left(\frac{4\Omega}{3}\right)^5\left(S_{22}\right)^2
\end{equation}
with
\begin{equation}
S_{22}=\sqrt{2}\frac{32\pi}{15}\frac{GM}{c^5}\alpha\Omega R^3 \widetilde{J} \nonumber
\end{equation}
 giving
\begin{equation}\label{djdt1}
\frac{dJ}{dt}=-\frac{2^{17}\pi}{3^7 5^2}\frac{G M^2}{c^7}\alpha^2\Omega^7 R^6 \widetilde{J}^2
\end{equation}

Again, the angular momentum loss couples to the emission of gravitational waves as\cite{thorne_rev_mod80},
\begin{equation}\label{djdt2}
\frac{dJ}{dt}=\frac{D^2 m \omega c^3}{4G}<h_+^2(t)+h_{\times}^2(t)>
\end{equation}
where $h_+(t)$ and $h_{\times}(t)$ are the strain amplitudes for the two polarisations of the gravitational wave, $D$ is the distance to the source and $<\cdots>$ denotes an average over possible orientations of the source with respect to the observer. Combining the Equations (\ref{djdt1}) and (\ref{djdt2}) we arrive at
\begin{equation}\label{basic_h}
<h_+^2(t)+h_\times^2(t)>=\frac{2^{16}\pi}{3^6 5^2}\frac{G^2 M^2 R^6 \widetilde{J}^2 \alpha^2}{c^{10} D^2}\Omega^6
\end{equation}

In the stationary phase approximation (which is always valid for a secular instability), this expression for the gravitational wave is written in terms of the Fourier transformed quantity as
\begin{equation}\label{hfourier_bracket}
<\widetilde{h}_+^2(f)+\widetilde{h}_\times^2(f)>=\frac{2^{16}\pi}{3^6 5^2}\frac{G^2 M^2 R^6 \widetilde{J}^2 \alpha^2}{c^{10} D^2}\Omega^6 \Big|\frac{df}{dt}\Big|^{-1}
\end{equation}

The quantity $\Big| \frac{df}{dt}\Big|$ can be calculated from the expression for $\frac{d\Omega}{dt}$ given by the Equation (\ref{domgdtfnl}) and using Equations (\ref{tauV}), (\ref{tauGRfnl}) and (\ref{tauM}) we get

\begin{eqnarray}
\frac{d\Omega}{dt} & = &
 \frac{8\alpha^2\Omega Q (K+2)}{3\tau_{GR}}
+\frac{2(4K+5)\alpha^2\Omega Q}{3\tau_V}
-\frac{\Omega}{\tau_M}  \nonumber \\
-\frac{3\pi}{2}\frac{df}{dt} & = & \frac{8\alpha^2 Q \Omega_0 (K+2)}{3\widetilde{\tau}_{GR}}\Omega_t^7-\frac{B_{14}^2 \Omega_0}{\widetilde{\tau}_M}\Omega_t^3 \nonumber \\
\Big|\frac{df}{dt}\Big|& = &\Big|\frac{16(K+2)\alpha^2 Q \Omega_0}{9\pi\widetilde{\tau}_{GR}}\Omega_t^7 - \frac{2\Omega_0 B_{14}^2}{3\pi \widetilde{\tau}_M}\Omega_t^3 \Big| \label{dfdt}
\end{eqnarray}

Here $\Omega_t=\Omega/\Omega_0$ with $\Omega_0=\sqrt{\pi \bar{\rho} G},\quad \bar{\rho}$ being the average density of a typical neutron star. Also, since the viscous timescale $\widetilde{\tau}_V$ compares very long (of the order of few years) with respect to either $\widetilde{\tau}_{GR}$ or $\widetilde{\tau}_M$ (of the order of few hours), we have neglected the viscosity term from the above equations.

With the help of Equation(\ref{dfdt}), it is now easy to rewrite Equation (\ref{hfourier_bracket}) as
\begin{widetext}
\begin{equation}\label{hf_no_angle_avg}
<\widetilde{h}_+^2(f)+\widetilde{h}_\times^2(f)>=\frac{2^{15}\pi}{3^5\; 5^2 }\frac{G^2 M^2 R^6 \widetilde{J}^2 \alpha^2 \Omega_0^5}{c^{10} D^2}\Omega_t^3   \;     \Big|\frac{8(K+2)\alpha^2 Q \Omega_t^4}{3\widetilde{\tau}_{GR}} - \frac{B_{14}^2}{\widetilde{\tau}_M} \Big|^{-1}
\end{equation}
\end{widetext}

The measured value of $\Big | \widetilde{h}(f) \Big|^2 $ depends on the location of the source in the observer's sky and the polarisation angle with respect to the interferometer. Since the gravitational wave interferometers are omnidirectional, we must have to effect an averaging over these angles to have a correct description of the measured strain values. Thus, averaged over these angles, the measured strain in the Fourier domain is given by,

\begin{equation}
<|\widetilde{h}(f)|^2>=\frac{1}{5}<|\widetilde{h}_+(f)|^2+<|\widetilde{h}_\times (f)|^2>
\end{equation}

Also, the average over the source location in three-dimensional space should also be taken into account because of the fact that spatial average weights more strongly those orientations giving out stronger signals and the effect of this averaging enhances $<|\widetilde{h}(f)|^2>$ further by about 1.5 \cite{thorne300yrs}. Combining these results, the average value of $\widetilde{h}(f)$ generated in a neutron star like source reduces to $3/10$-th of that given in Equation (\ref{hf_no_angle_avg}) and we have,

\begin{widetext}
\begin{equation}
\widetilde{h}(f) \sim \frac{64}{75}\,\sqrt{\frac{2\pi}{3}}\, \frac{GMR^3\widetilde{J}\alpha\Omega_0^{5/2}}{c^5\,D}\Omega_t^{3/2}\Bigg|\frac{8}{3}\frac{(K+2)\alpha^2 Q\Omega_t^4}{\widetilde{\tau}_{GR}}-\frac{B_{14}^2}{\widetilde{\tau}_M}\Bigg|^{-1/2}
\end{equation}
\end{widetext}

For a typical neutron star of mass $M=1.4M_\odot$, and radius $R=12.53$ km emitting an average value of $\widetilde{h}$ at a distance of $D=20$ Mpc is then given by,

\begin{equation}\label{hf_final}
\widetilde{h}(f)=3.3\times 10^{-26} \left[ \frac{\Omega_t (K+2)}{13}+\frac{B_{14}^2}{6.74\times 10^4 \,\alpha^2 \Omega_t^3} \right]^{-1/2}
\end{equation}

The characteristic gravitational wave amplitude as a function of time can be found using Equation (\ref{basic_h}). When averaged over source and detector orientation, it is given by
\begin{equation}
\big|h(t)\big|=6.6\times 10^{-24}\alpha(t)\left(\frac{\Omega(t)}{\Omega_0}\right)^3\left(\frac{20 \mbox{Mpc}}{D}\right),
\end{equation}

where $D$ is the distance to the source in 20 Mpc and $\Omega_0=\sqrt{\pi \bar{\rho} G}$. We have considered the evolution for a period of one year from the inception of the instability. Figures (\ref{gw_time_b0_k.lbl}) through (\ref{gw_time_b_ke12.lbl}) depict the evolution of absolute value of gravitational wave strain during $r$-mode instability for various values of magnetic fields B and the differential rotation parameter K. During the early part of the evolution, that is, upto about 12 minutes in case of K=0 and about 4 minutes for K=$10^{12}$, $|h(t)|$ grows exponentially while during the later part, extending upto 1 year, it decreases slowly. At its maximum, $h(t)$ assumes a value of $5\times 10^{-25}$ for K=0 and $1\times 10^{-30}$ for K=$10^{12}$, both for B$_{14}$=0. The cases with other values of $K$ and $B_{14}$ also behave similarly.

It is worth mentioning here that in Ref. \cite{OLCSVA} an expression for the frequency-domain gravitational wave amplitude $|\widetilde{h}(f)|$ was derived based on the assumption that $dJ/df \varpropto I$, where $J$ is the total angular momentum of the star, $f$ is the frequency of the emitted gravitational wave and $I$ is the moment of inertia of the unperturbed star. Since within the model of Ref. \cite{OLCSVA}, the condition $dJ/df \varpropto I$ only applies during the second stage of evolution, the expression obtained there for $|\widetilde{h}(f)|$ is valid just for this stage of evolution. Within our consideration it can be shown that the condition  $dJ/df \varpropto I$ easily holds, not only during the secnd stage of evolution, but also during the first stage. 

\section{Analysis of Results}
First, we choose to consider the duration of evolution of $l=2$ $r$-mode instability to be a period of 1 year from the inception of the instability. During this period, the  dissipative effects of bulk viscosity $\tau_B$ is hardly dominant to damp the mode. Temperature during this period falls from about $10^{11}$ K to $10^9$ K (Equation (\ref{temp_time})). We wish to stop the evolution beyond 1 year since temperature achieved beyond 1 year of evolution, superfluid effects are expected to become important, rendering invalid our assumptions about the viscous timescale \cite{lind_mend95}.

The amplitude of the $r$-mode at the inception of the instability is taken to be as small as $10^{-8}$. The problem of solving the Equations (\ref{daldtfnl}) and (\ref{domgdtfnl}) with no contribution from $\tau_B$ then is amenable to straight numerical solution by Runge-Kutta 4-th order method.

The curves for growth of $r$-mode amplitude $\alpha$, the evolution of angular velocity $\Omega$ and the gravitational wave strain amplitude $h$, suggest that the entire $r$-mode instability growth proceeds in two stages - the first stage and the second stage. During the first stage, the early stage, which lasts about few hundred seconds, the mode amplitude and gravitational wave strain amplitude exponentially rise and the angular velocity maintains more or less a constant value. During the second stage, the later stage after the first stage, lasting for the rest period of 1 year, the angular velocity falls abruptly. The mode amplitude at this stage gets saturated to maintain a roughly constant value or register a very slow rise. The gravitational wave at the second stage falls to a very low value by the end of 1 year.

Several observations in the results are in order.


The amplitude of the $r$-mode first rises exponentially and saturates in a natural way a few hundred seconds after the mode instability sets in. This is unlike the case of \cite{OLCSVA}, where the first stage is to be terminated by hand and for second stage a different set of equations need evolving.

For a small initial differential rotation ($K\simeq 0$), the $r$-mode saturates at values of the order of unity. On the other hand, for significantly larger values of $K$, \emph{i.e.}, $K \gg 1$,  the saturation amplitude can become as small as $10^{-5}$ when magnetic field is absent. With the presence of magnetic field it can go down further to $10^{-7}$. Also the differential rotation parameter  $K$ causes to saturate the growth earlier than the one with $K=0$.

Higher values of magnetic field ($B_{14}$=6, 10, 15 and 20, for example) causes the growth of mode amplitude to suppress drastically. The growth curves for $\alpha$ labelled with different $K$ are merged at these higher values of magnetic fields (Figures \ref{alpha_time_b0_k.lbl} upto \ref{alpha_time_b20_k.lbl}).

\begin{figure}[htp]
\includegraphics[width=0.5\textwidth,height=0.3\textheight]{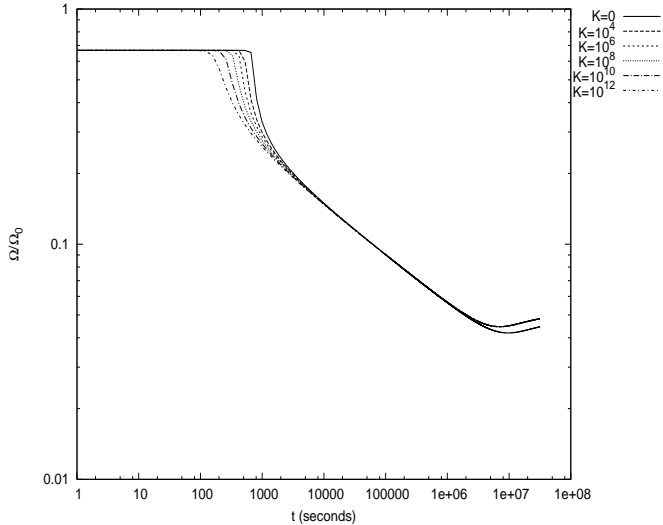}
\caption{Angular velocity evolution for different K and $B_{14}=0$.}\label{omega_time_b0_k.lbl}
\end{figure}

\begin{figure}[htp]
\includegraphics[width=0.5\textwidth,height=0.3\textheight]{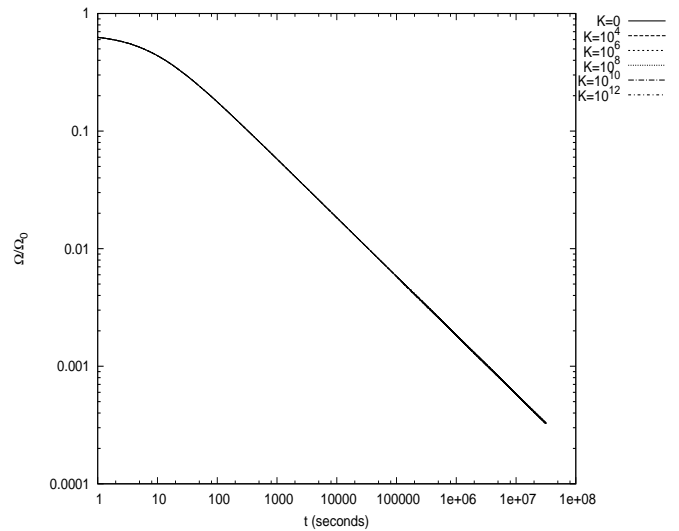}
\caption{omega vs time for different k and $ B_{14}=100 $ }\label{omega_time_b100_k.lbl}
\end{figure}

\begin{figure}[htp]
\includegraphics[width=0.5\textwidth,height=0.3\textheight]{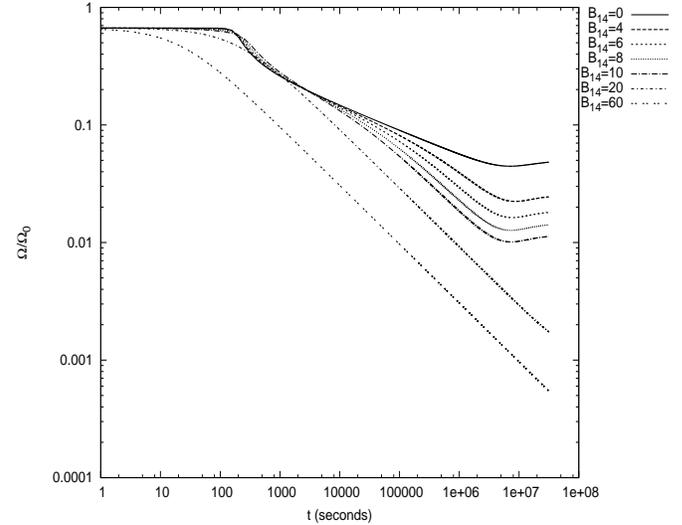}
\caption{omega vs time for different B and $ K=10^{12} $ }\label{omega_time_b_ke12.lbl}
\end{figure}

The angular velocity of the star remains approximately constant during the first stage of evolution and abruptly falls to a very low value in the next few thousand seconds after the first stage is over. This is true for all $K$ and $B_{14}$. However, there exist minor differences in the angular velocity evolution curves with various $K$ and $B_{14}$ values. Essentially, the abrupt falling of the angular velocity takes place earlier for higher $K$ values (Figure \ref{omega_time_b0_k.lbl} through \ref{omega_time_b_ke12.lbl}).  Additionally, distinct trajectories are found at early part of the second stage in case of larger values of $K$ with a moderately high value of magnetic field. Thereafter,  the curves approach each other to become very closely spaced, signifying a final angular velocity with more or less a single value at the end of the second stage, \emph{i.e.,} at the end of 1 year.

For sufficiently high values of magnetic fields ($B_{14}$=20 or above), the distinct curves associated with different values of $K$ all merged to a single curve. This signifies the fact that the effect of differential rotation on the evolution of $r$-mode instability can be erased by a large magnetic field.

\begin{figure}[htp]
\includegraphics[width=0.5\textwidth,height=0.3\textheight]{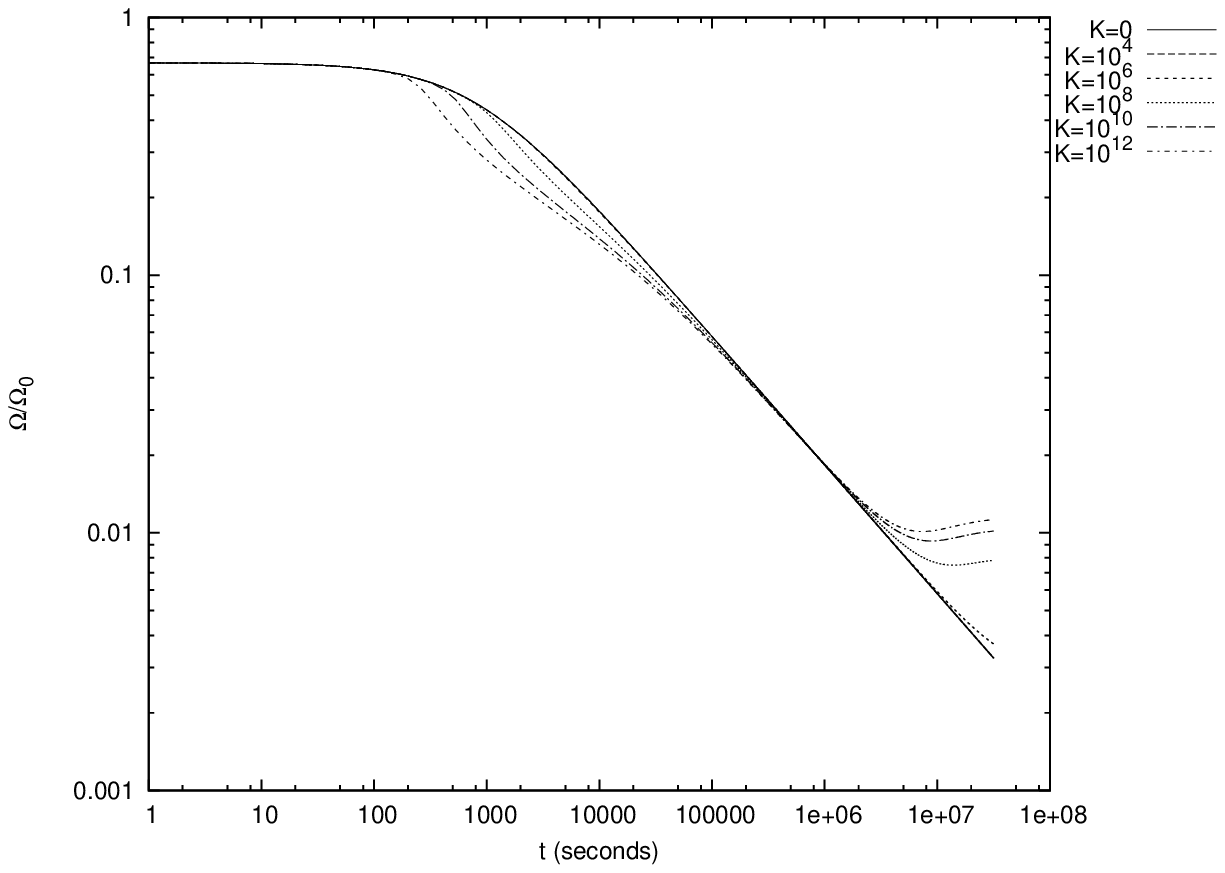}
\caption{omega vs time for different k and $ B_{14}=10 $ }
\end{figure} 

\begin{figure}[htp]
\includegraphics[width=0.5\textwidth,height=0.3\textheight]{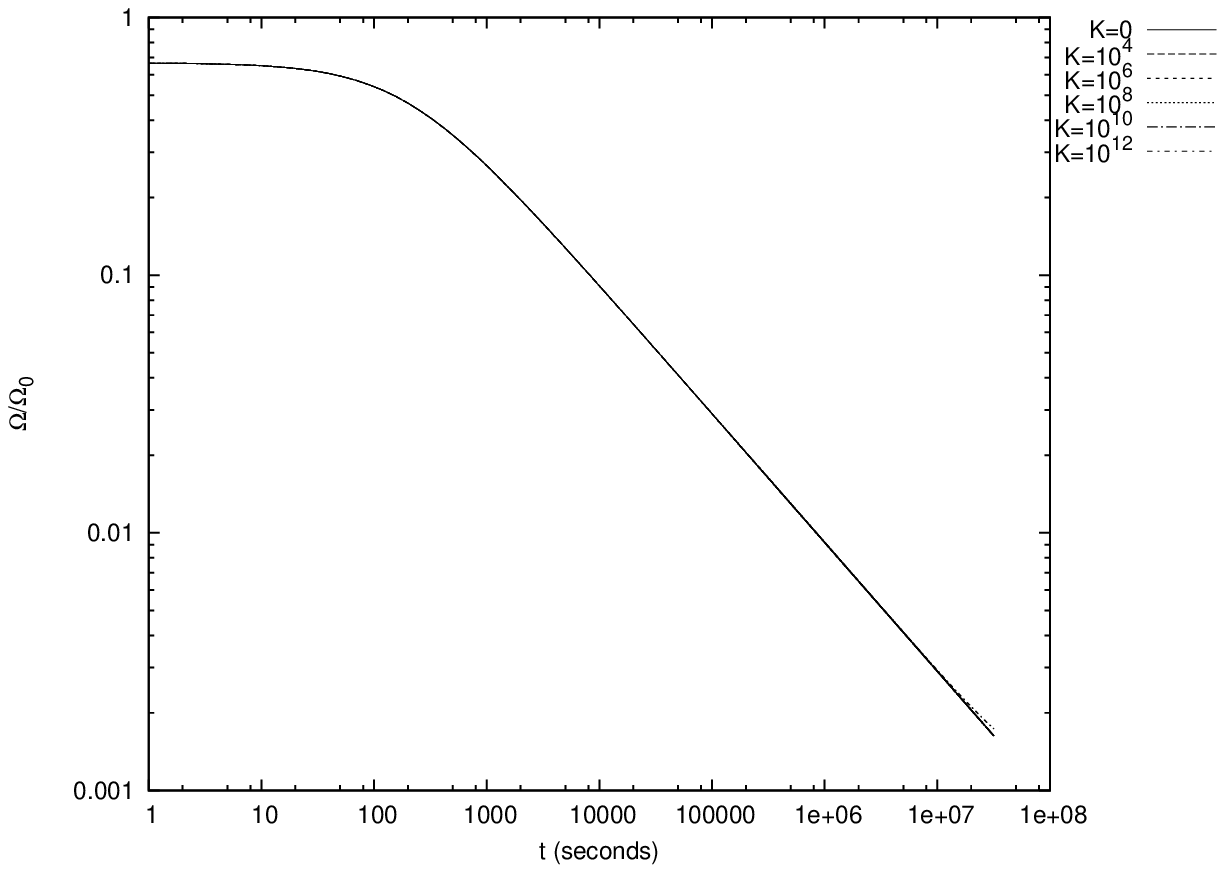}
\caption{omega vs time for different k and $ B_{14}=20 $ }
\end{figure} 

\begin{figure}[htp]
\includegraphics[width=0.5\textwidth,height=0.3\textheight]{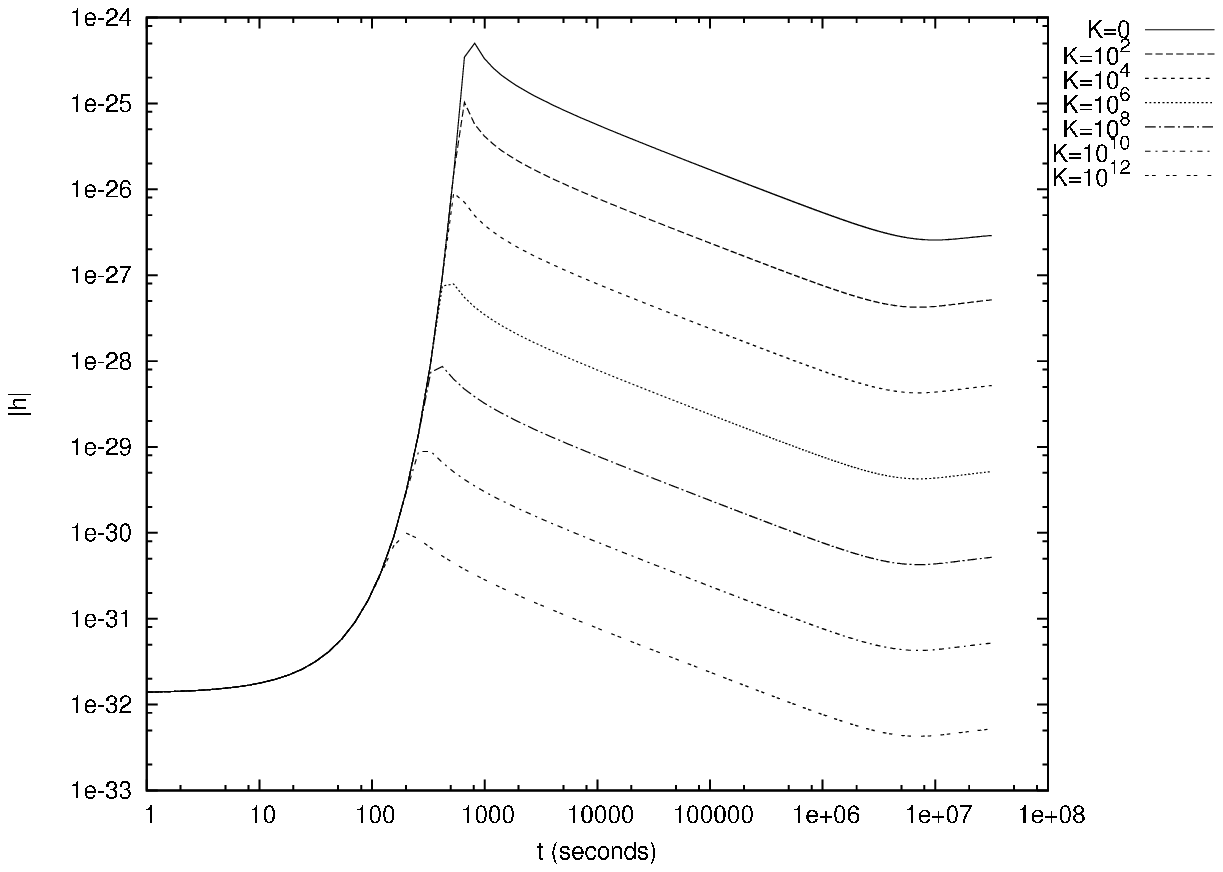}
\caption{Gravitational wave amplitude vs time for different k and $ B_{14}=0 $ }\label{gw_time_b0_k.lbl}
\end{figure}

\begin{figure}[htp]
\includegraphics[width=0.5\textwidth,height=0.3\textheight]{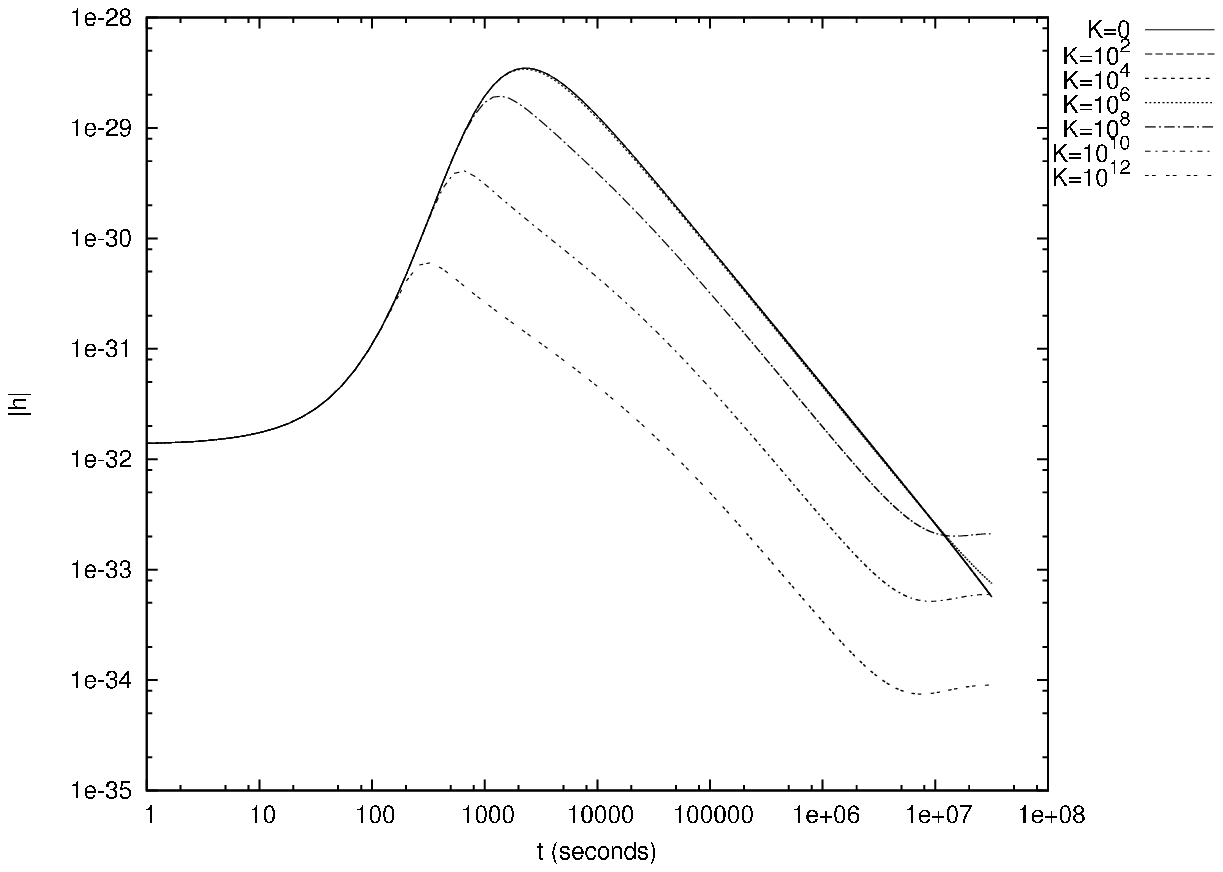}
\caption{Gravitational wave amplitude vs time for different k and $ B_{14}=10 $ }\label{gw_time_b10_k.lbl}
\end{figure}

\begin{figure}[htp]
\includegraphics[width=0.5\textwidth,height=0.3\textheight]{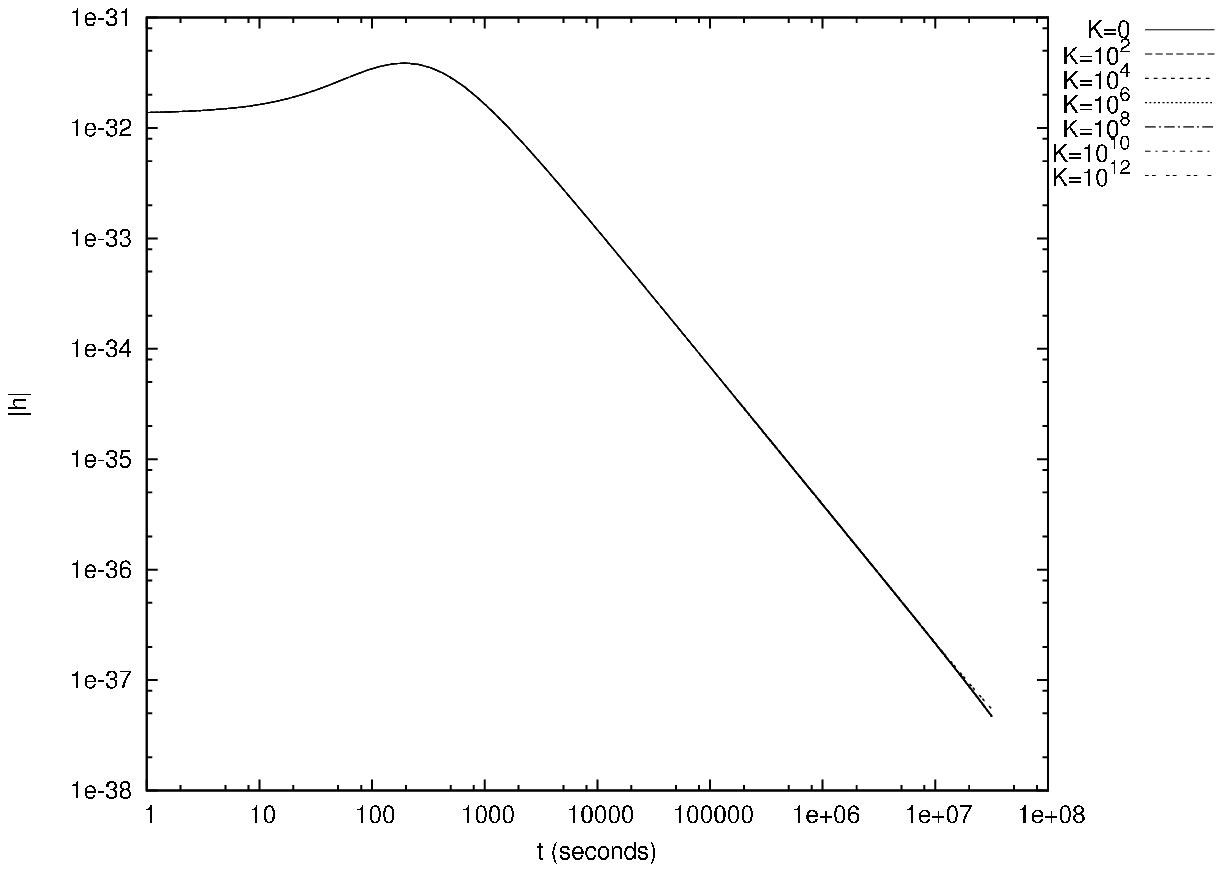}
\caption{Gravitational wave amplitude vs time for different k and $ B_{14}=20 $ }\label{gw_time_b20_k.lbl}
\end{figure}

\begin{figure}[htp]
\includegraphics[width=0.5\textwidth,height=0.3\textheight]{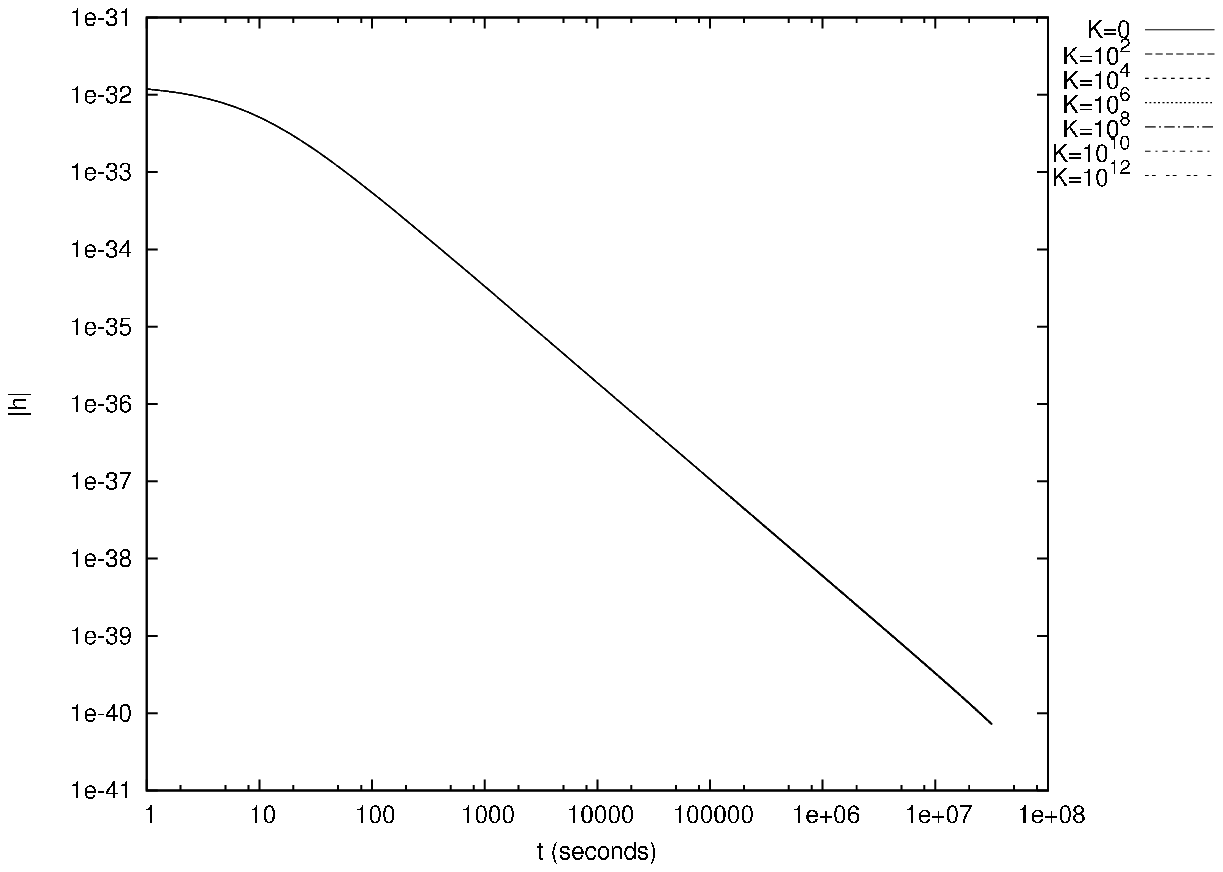}
\caption{Gravitational wave amplitude vs time for different k and $ B_{14}=100 $ }\label{gw_time_b100_k.lbl}
\end{figure}

\begin{figure}[htp]
\includegraphics[width=0.5\textwidth,height=0.3\textheight]{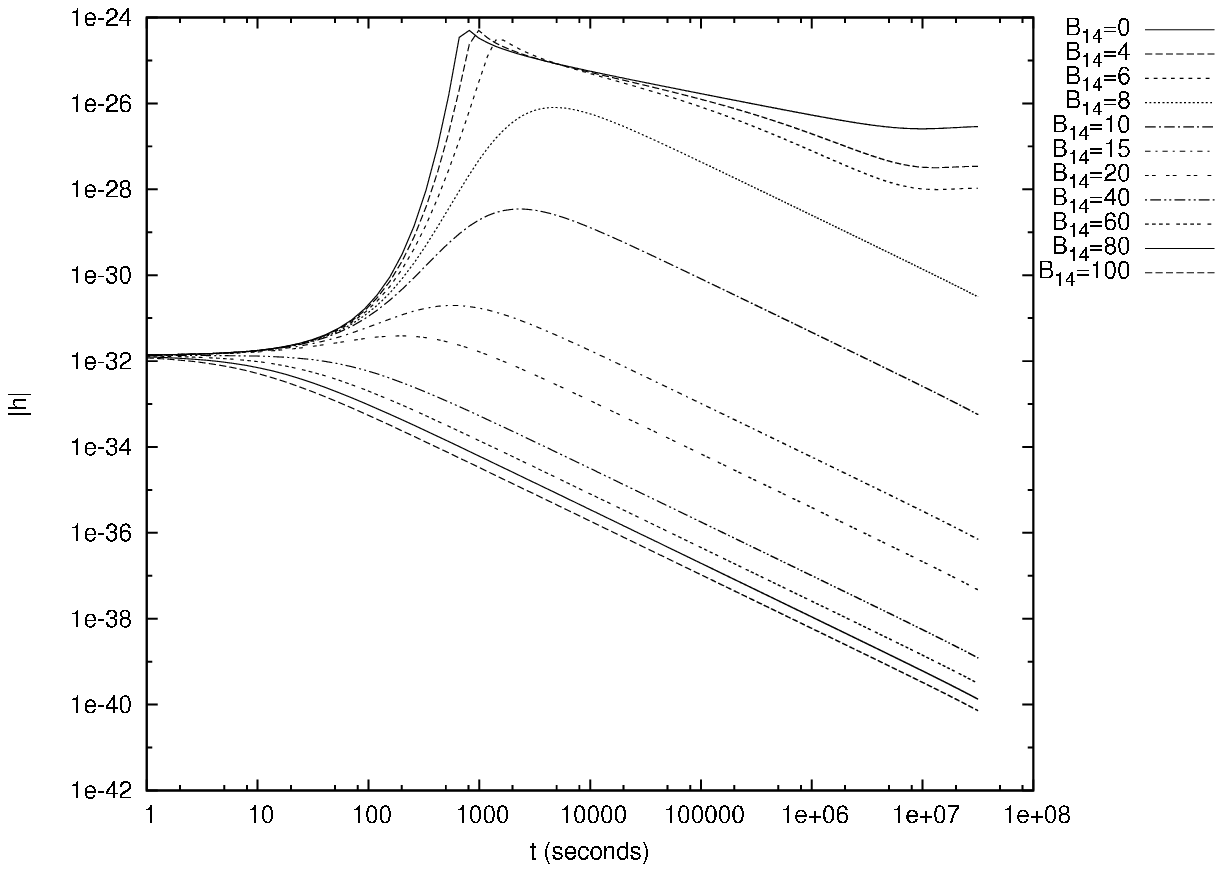}
\caption{Gravitational wave amplitude vs time for different strengths of magnetic field and the differential rotation parameter $K=0$ }\label{gw_time_b_k0.lbl}
\end{figure}

\begin{figure}[htp]
\includegraphics[width=0.5\textwidth,height=0.3\textheight]{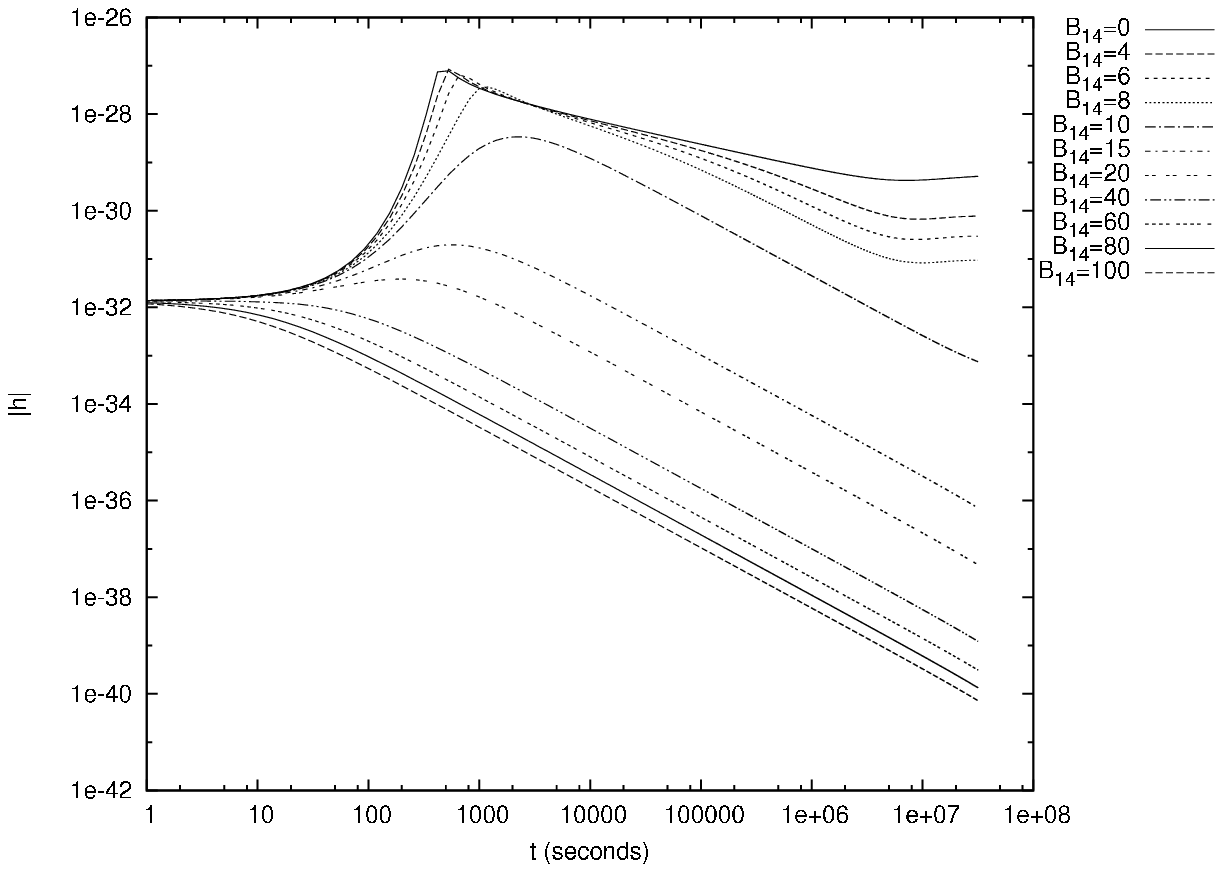}
\caption{Gravitational wave amplitude vs time for different strengths of magnetic field and the differential rotation parameter $K=10^6$ }\label{gw_time_b_ke06.lbl}
\end{figure}

\begin{figure}[htp]
\includegraphics[width=0.5\textwidth,height=0.3\textheight]{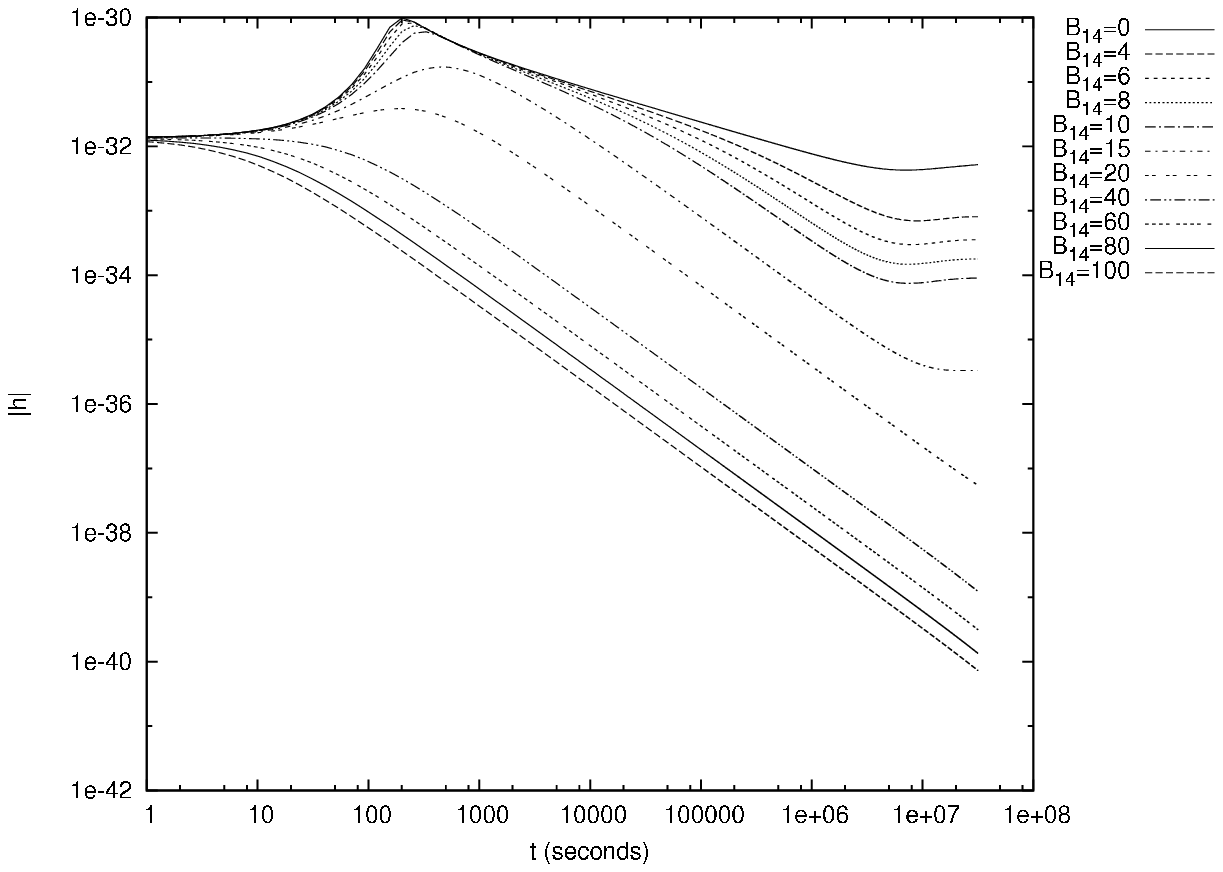}
\caption{Gravitational wave amplitude vs time for different strengths of magnetic field and the differential rotation parameter $K=10^{12}$ }\label{gw_time_b_ke12.lbl}
\end{figure}

For investigating the gravitational wave emitted from the $r$-mode instability, we choose the same set of parameters as in the earlier section to start the numerical solutions of the Equations (\ref{daldtfnl}) and (\ref{domgdtfnl}) and estimate the gravitational wave.  Figures (\ref{gw_time_b0_k.lbl}) through (\ref{gw_time_b_ke12.lbl} depict the results in graphical form.

Analysis of the graphs reveals that the amplitude of the gravitational wave grows exponentially within first few hundred seconds (first stage) to reach its maximum after which it falls gradually in approximately linear fashion. As the initial differential rotation parameter $K$ increases, the maxima of the gravitational wave strain amplitude is pushed towards less time, \emph{i.e.},the maxima reaches earlier and earlier moments as $K$ increases. At the same time the magnitude of the maxima also becomes lesser. As for example, the gravitational wave amplitude reaches its maximum of $h_{max}\sim 5\times 10^{-24}$ for $K=0$ and $h_{max}\sim 10^{-30}$ for $K=10^{12}$ (Figure (\ref{gw_time_b0_k.lbl})) for the magnetic field $B_{14}=0$. Larger values of magnetic field renders $h_{max}$ to suppress to very low magnitude. Also observed is the influence of magnetic field to cause the turning point at the maxima to be more and more smoothened unlike the case of no magnetic field where the turning point has an abrupt change. Additionally, larger and larger magnetic fields tend to erase the distinction between the curves with different values of initial differential rotation parameter $K$ and merge all the curves to register a very low value of  $h_{max}<10^{-30}$ for $ B_{14}=20 $. The effects of magnetic fields for different $K$ on gravitational wave output may be clear from the graphs in Figures (\ref{gw_time_b_k0.lbl}) through (\ref{gw_time_b_ke12.lbl}). Similarly the effects of the differential rotation factor for different magnetic field strengths are depticted in Figures (\ref{gw_time_b0_k.lbl}) through (\ref{gw_time_b20_k.lbl}).


It is worth mentioning here that within the model of evolution proposed in Ref. \cite{OLCSVA}, the frequency-domain gravitational wave amplitude has a spike at high frequencies, due to the fact that during the first stage of evolution the angular velocity of the star evolves very slowly on the viscous timescale, leading to a quasi-monochromatic gravitational wave emission during the first 500 seconds of evolution. However, as we have seen, if one takes into account the influence of differential rotation, namely, the fact that it contributes to the physical angular momentum of the $r$-mode perturbation, then the angular velocity of the star evolves in the gravitational-radiation timescale already in the first stage of evolution. Influence of the magnetic field is merely to change the overall structure of the evolution because of the fact that the magnetic field term appears in the evolution equations (\ref{daldtfnl}) and (\ref{domgdtfnl}) as a separate additive term. As a consequence, the gravitational wave amplitude $|\widetilde{h}(f)|$ in the first stage of evolution is also given by Equation (\ref{hf_final}) and we observe no spike.

\begin{figure}[htp]
\includegraphics[width=0.5\textwidth,height=0.3\textheight]{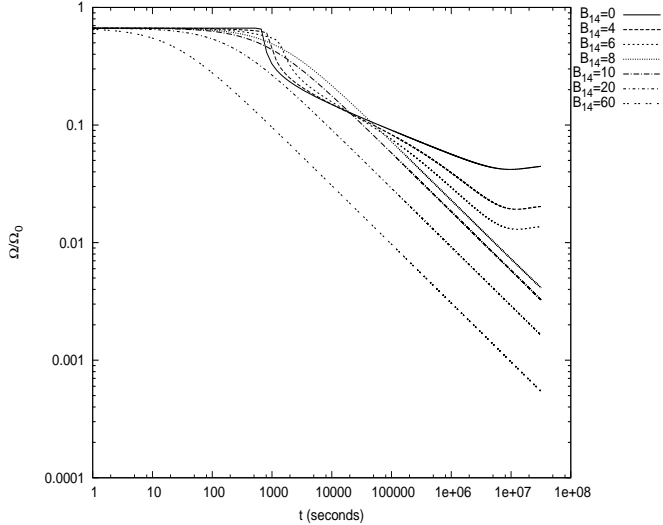}
\caption{omega vs time for different B and $ K=0 $ }
\end{figure} 


\begin{figure}[htp]
\includegraphics[width=0.5\textwidth,height=0.3\textheight]{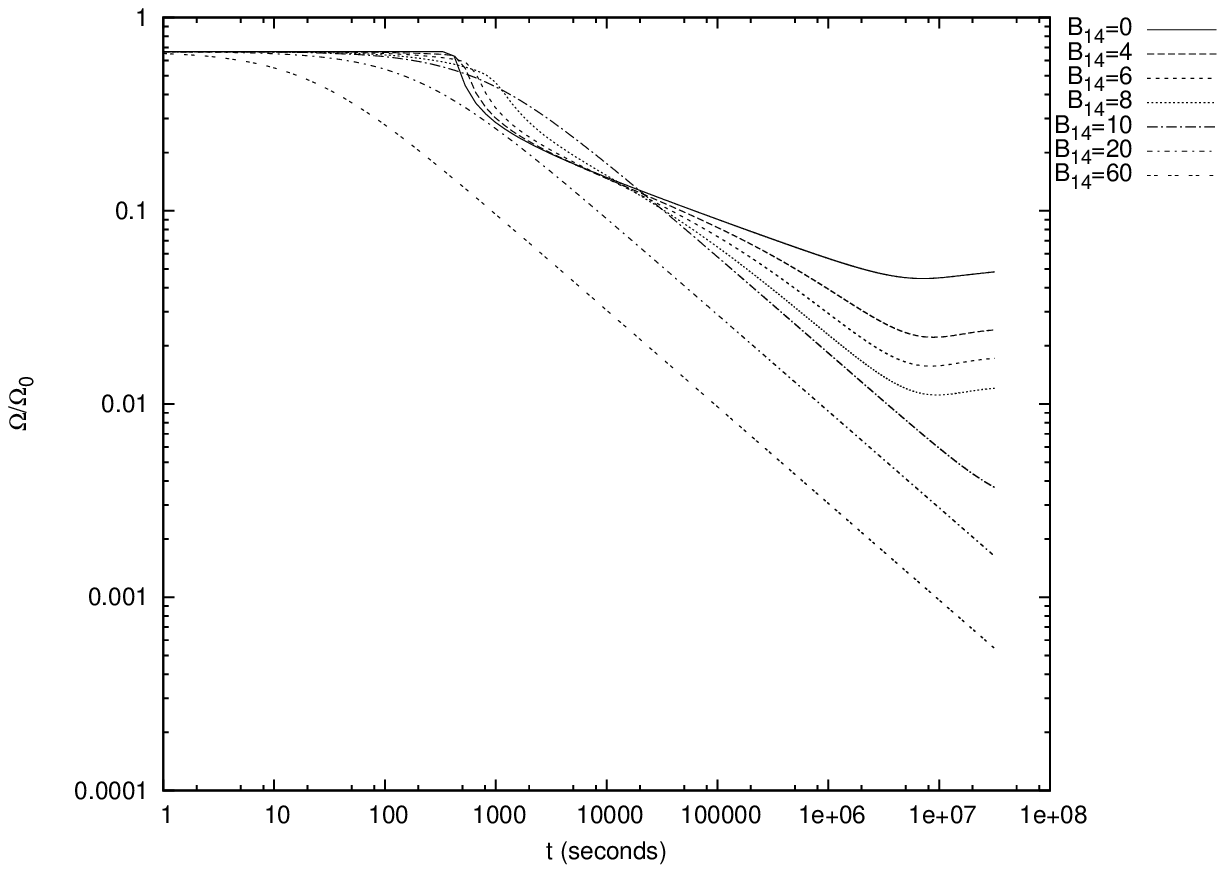}
\caption{omega vs time for different B and $ K=10^6 $ }
\end{figure} 


\section{Detection of Gravitational Waves from $r$-mode Instability in the presence of Magnetic Field}
The gravitational radiation emitted by a newly born neutron star, as is spins down due to the $r$-mode instability, could in principle be detected by the laser interferometric detectors. In the following we discuss the detectibility of gravitational waves during the spin-down of neutron star by detectors like LIGO, Advance LIGO and VIRGO, using appropriate detecting strategies.

In reality,  neutron stars are such a complex system involving many other effects such as superfluidity, quark-gluon plasma, a very complicated magnetic phenomena, our knowledge of the evolution of the $r$-mode instability is insuffecient to predict the gravitational waveform with such an accuracy that a matched filtering signal-processing technique would be feasible. Therefore in this work, the match filtering is used only to estimate the detectibility of the gravitational-wave signal for the models parameters we have considered here. For this purpose, we construct the characteristic amplitude of the signal through the relation

\begin{widetext}
\begin{eqnarray}
h_c(f)&\equiv& f\big|\widetilde{h}(f)\big|  \nonumber\\
&=& 3.3\times 10^{-26}f\left[\frac{\left(f/f_t\right)(K+2)}{13}+\frac{B_{14}^2}{6.74\times 10^4\,\alpha^2\,\left(f/f_t\right)^3}\right]^{-1/2}
\end{eqnarray}
\end{widetext}
 
is compared with the rms strain noise in the detector,
\begin{equation}
h_{rms}(f)\equiv \sqrt{fS_h(h)},
\end{equation}
where $S_h(f)$ is the noise power spectral density of the detector.
For frequencies in the interval in the interval of 50$\lesssim f \lesssim$ 1200 Hz, the curves for the noise power spectral densities of the LIGO \cite{ligo_in}, advanced LIGO \cite{ligo_adv} and VIRGO \cite{virgo} detectors are well approximated by the following analytical expressions, respectively:

\begin{widetext}
\begin{equation}
S_h(h) = S_1 \left[\left(\frac{f_1}{f}\right)^4+\left(\frac{f}{f_1}\right)^2\right],
\end{equation}
where $S_1=3.4\times 10^{-46}\mbox{Hz}^{-1}$ and $f_1$=142.0 Hz,
\begin{equation}
S_h(f)=S_2\Bigg\{1+\left(\frac{f_2}{f}\right)^7 -\frac{10}{3}\left(\frac{f}{f_3}\right)\left[1-\left(\frac{f}{f_3}\right)+\frac{3}{50}\left(\frac{f}{f_3}\right)^2\right]\Bigg\},
\end{equation}
where $S_2=2.2\times 10^{-47} \mbox{Hz}^{-1}, \;\; f_2=52.8 \mbox{Hz and } f_3=421.3$ Hz, and
\begin{equation}
S_h(f)=S_3\left[ 1 + \frac{1}{6}\Bigg\{\left(\frac{f_4}{f}\right)^2+\left(\frac{f}{f_4}\right)^2\Bigg\}\right] ,
\end{equation}
where $S_3=1.5\times 10^{-45} \mbox{Hz}^{-1},\;\; f_4=249.6$ Hz.
\end{widetext}

\begin{figure}[htp]
\includegraphics[width=0.5\textwidth,height=0.25\textheight]{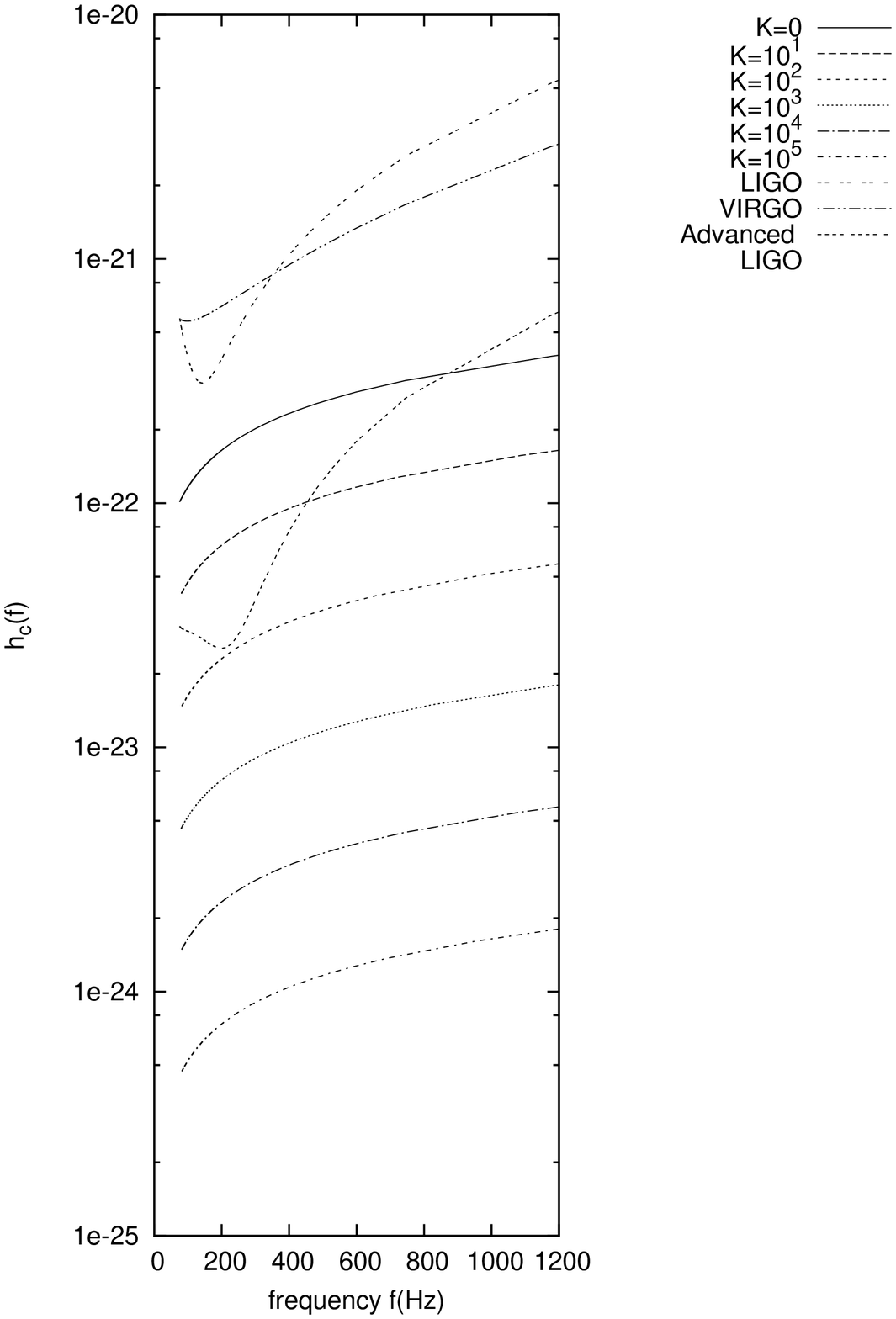}
\caption{Characteristic gravitational wave amplitude vs frequency for different strengths of the differential rotation parameter $K$ and magnetic field strength $B_{14}=0$.}\label{gwhc_freq_b0_k.lbl}
\end{figure}

\begin{figure}[htp]
\includegraphics[width=0.5\textwidth,height=0.25\textheight]{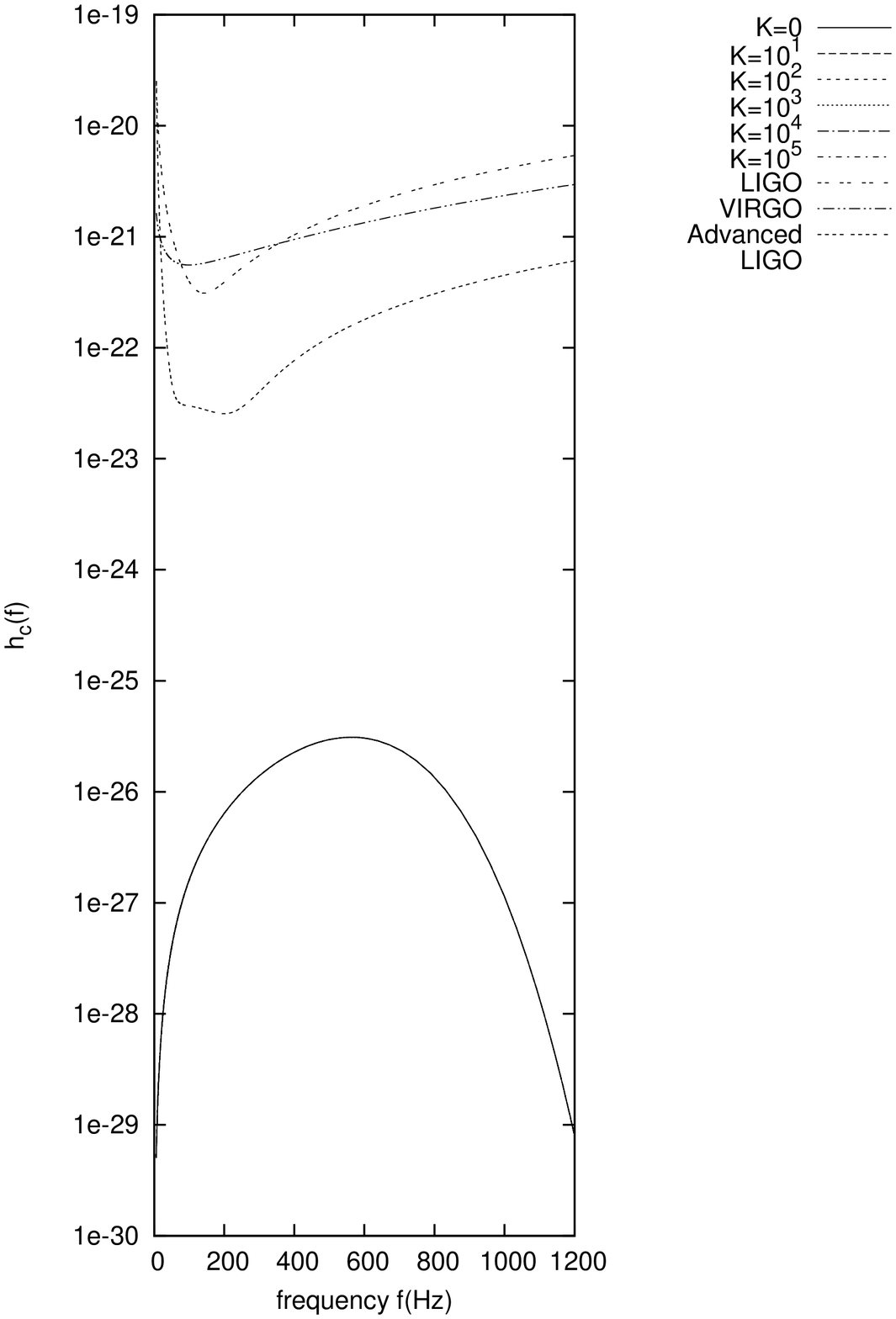}
\caption{Characteristic gravitational wave amplitude vs frequency for different strengths of the differential rotation parameter $K$ and magnetic field strength $B_{14}=10$.}\label{gwhc_freq_b10_k.lbl}
\end{figure}

\begin{figure}[htp]
\includegraphics[width=0.5\textwidth,height=0.25\textheight]{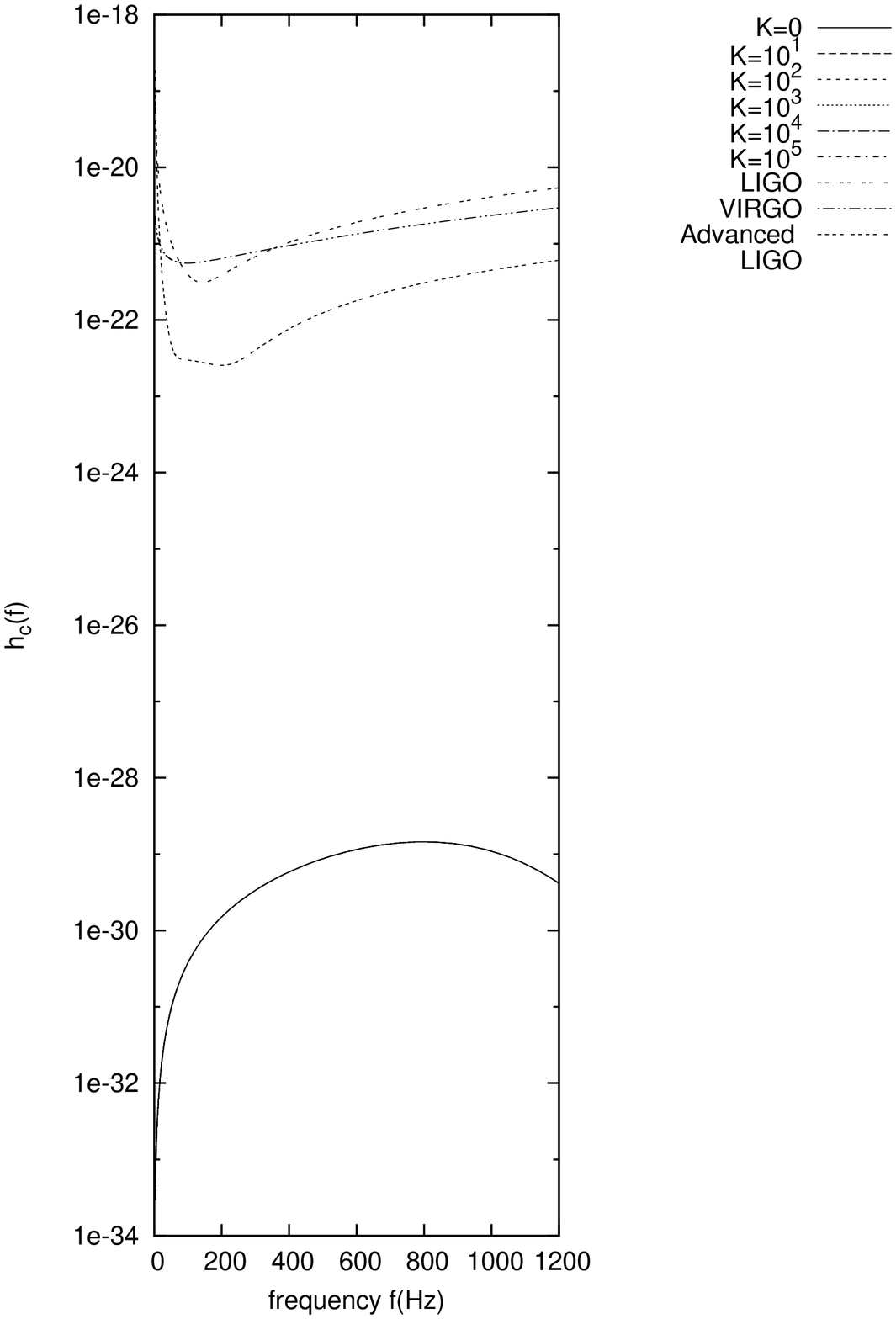}
\caption{Characteristic gravitational wave amplitude vs frequency for different strengths of the differential rotation parameter $K$ and magnetic field strength $B_{14}=20$.}\label{gwhc_freq_b20_k.lbl}
\end{figure}

\begin{figure}[htp]
\includegraphics[width=0.5\textwidth,height=0.3\textheight]{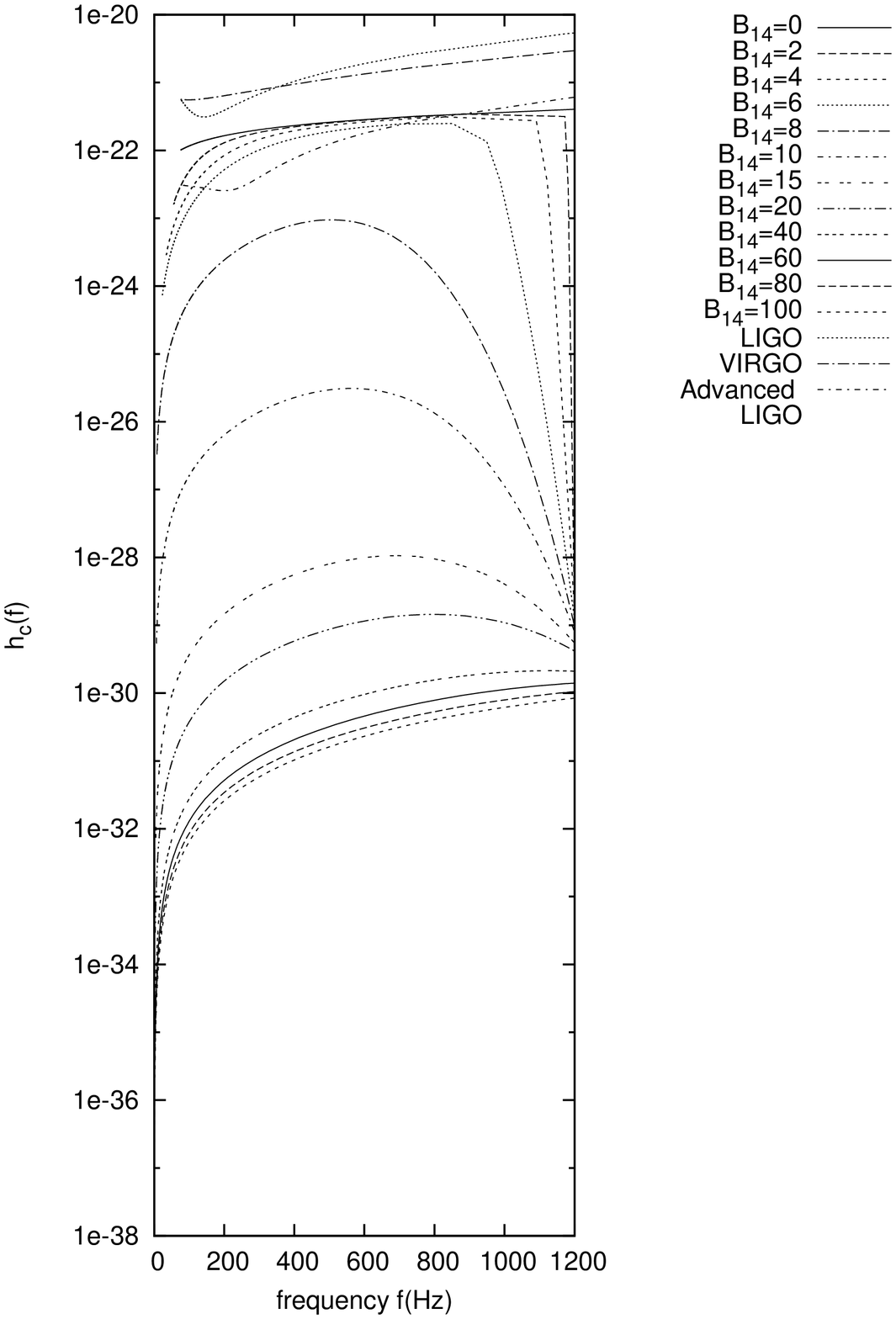}
\caption{Characteristic gravitational wave amplitude vs frequency for different strengths of the magnetic field and the differential rotation parameter $K=0$.}\label{gwhc_freq_b_k0.lbl}
\end{figure}

\begin{figure}[htp]
\includegraphics[width=0.5\textwidth,height=0.3\textheight]{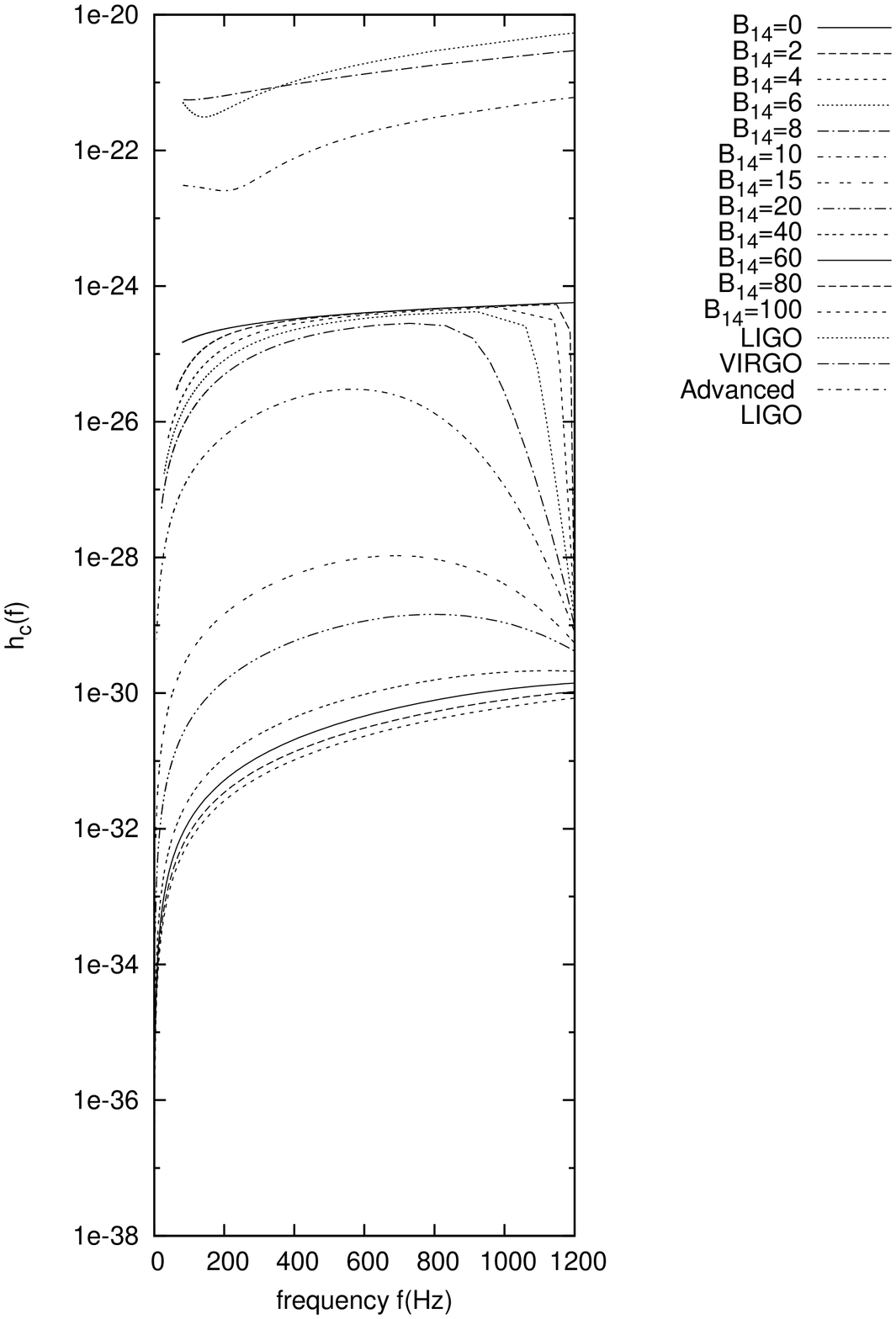}
\caption{Characteristic gravitational wave amplitude vs frequency for different strengths of the magnetic field and the differential rotation parameter $K=10^6$.}\label{gwhc_freq_b_ke06.lbl}
\end{figure}

\begin{figure}[htp]
\includegraphics[width=0.5\textwidth,height=0.3\textheight]{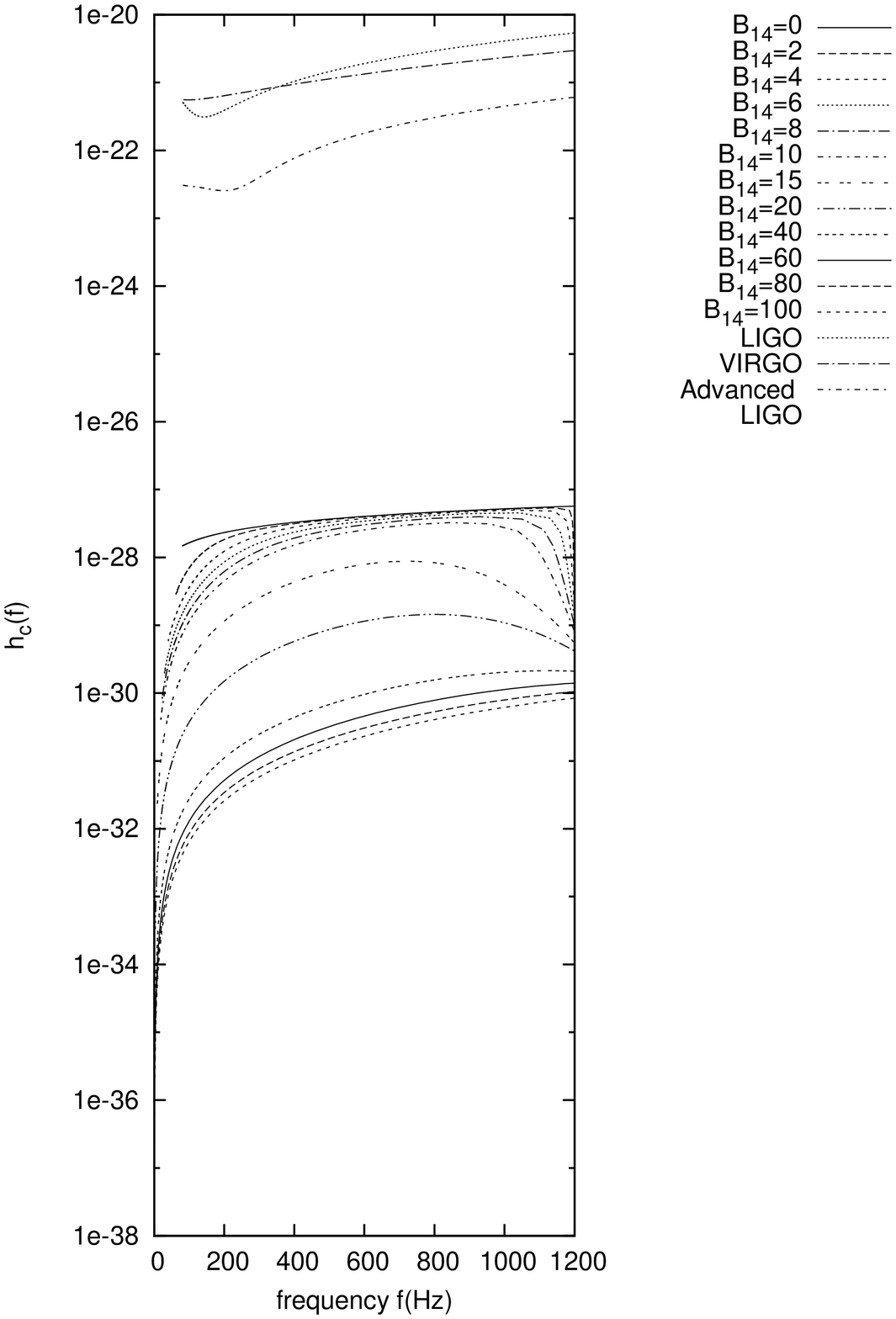}
\caption{Characteristic gravitational wave amplitude vs frequency for different strengths of the magnetic field and the differential rotation parameter $K=10^{12}$.}\label{gwhc_freq_b_ke12.lbl}
\end{figure}

In Figures (\ref{gwhc_freq_b0_k.lbl}) through (\ref{gwhc_freq_b_ke12.lbl}) the curves corresponding to the rms strain noises in the initial LIGO, VIRGO and advanced LIGO detectors are compared with the curves corresponding to the characteristic amplitude of the gravitational wave signal for different values of differential rotation parameter K, the magnetic field B and for a distance to the source of D=20Mpc. For such a distance, which includes the Virgo cluster of galaxies, a few supernovae per year is expected.

The most striking feature of these figures is that the detectibility of gravitational waves from the $r$-mode instability of newly born neutron stars is drastically reduced as the initial value of differential rotation associated with $r$-modes increases. Similarly the stronger magnetic fields also contribute to the reduction of the signal-to-noise ratio of the these signals. (Table \ref{snrout}).

The visual comparison in Figures (\ref{gwhc_freq_b0_k.lbl}) through (\ref{gwhc_freq_b_ke12.lbl}) between the charateristic amplitude of the signal and the rms strain noise in the detector gives us a measure of the signal-to-noise ratio for matched filtering. A quantitative determination of the signal-to-noise ratio is obtained from,
\begin{equation}\label{matchedfilter}
\left( \frac{S}{N}\right)^2=2 \int_{f_{min}}^{f_{max}} \frac{df}{f} \left(\frac{h_c}{h_{rms}}\right)^2 .
\end{equation}
We have chosen the minimum and maximum value of frequency, $f_{min}$=50 Hz and $f_{max}$=1200 Hz for estimating the signal-to-noise ratio here.

We have computed the Signal-to-Noise ratios for the gravitational waves emitted due to the $r$-mode instability in a typical Neutron Star of mass 1.4 $M_\odot$ and radius 12.53 km. for different values of the differential rotation parameter and dipole magnetic field strengths by using the standard formula (\ref{matchedfilter}) for matched filtering technique. The results are tabulated in the Table \ref{snrout}. For the initial LIGO and VIRGO detectors, even for small initial differential rotation (K$\approx$0), the signal-to-noise ratio is not significant for D= 20 Mpc. However, this ratio can be increased if we consider sources located at smaller distances, but that will be at the cost of decreasing the number of expected supernova events per year and, hence the probability of a detection.

For the advanced LIGO detector the situation improves: for small values of $K$, and B, the signal-to-noise ratio is significant even for D = 20 Mpc. Therefore, and since within such a distance several supernovae per year are expected, one could hope that advanced LIGO detectors would be able to detect gravitational radiation for the $r$-mode instability of young neutron stars.

\begin{table}
\caption[$r$-mode detectibility Table]{Signal-to-Noise ratio of gravitional waves due to $r$-mode instability calculated for different detectors}\label{snrout} 
\begin{ruledtabular}
\begin{tabular}{ccccc}
B$_{14}$ &K   &   SNR@LIGO &   SNR@Adv. LIGO & SNR@VIRGO \\
\hline
  0.0  &  0.0 &   0.66 &   9.26 &  0.50\\
  2.0  &  0.0 &   0.53 &   7.57 &  0.43\\
  4.0  &  0.0 &   0.39 &   5.67 &  0.35\\
  6.0  &  0.0 &   0.28 &   4.02 &  0.27\\
  8.0  &  0.0 &   0.01 &   0.21 &  0.01\\
  10.0  &  0.0 &  0.00 &  0.008 &   0.00\\
&&&&\\
  0.0 &   10 &   0.27 &   3.75 &   0.20\\
  2.0 &   10 &   0.21 &   3.07 &   0.17\\
  4.0 &   10 &   0.16 &   2.32 &   0.15\\
  6.0 &   10 &   0.12 &   1.67 &   0.11\\
  8.0 &   10 &   0.02 &   0.20 &   0.01\\
  10.0 &   10 &  0.00 &   0.00 &   0.00\\
&&&&\\
  0.0 &   100 &   0.09 &   1.28 &   0.07\\
  2.0 &   100 &   0.07 &   1.06 &   0.06\\
  4.0 &   100 &   0.06 &   0.80 &   0.05\\
  6.0 &   100 &   0.04 &   0.58 &   0.04\\
  8.0 &   100 &   0.01 &   0.16 &   0.01\\
  10.0 &   100 &  0.00 &   0.00 &   0.00\\
&&&&\\
  0.0 &   1000 &   0.03 &   0.41 &   0.020\\
  2.0 &   1000 &   0.02 &   0.34 &   0.019\\
  4.0 &   1000 &   0.018 &   0.26 &   0.015\\
  6.0 &   1000 &   0.013 &   0.19 &   0.013\\
  8.0 &   1000 &   0.006 &   0.086 &   0.006\\
  10.0 &   1000 &   0.000 &   0.001 &   0.000\\
&&&&\\
  0.0 &   $10^4$ &   0.009 &   0.13 &   0.007\\
  2.0 &   $10^4$ &   0.007 &   0.11 &   0.006\\
  4.0 &   $10^4$ &   0.006 &   0.08 &   0.005\\
  6.0 &   $10^4$ &   0.004 &   0.06 &   0.004\\
  8.0 &   $10^4$ &   0.002 &   0.03 &   0.002\\
  10.0 &   $10^4$ &   0.000 &   0.001 &   0.000\\

\end{tabular}
\end{ruledtabular}
\end{table}

\section{Conclusions}
We have evaluated the behaviour of $r$-mode instability including its gravitational wave detectability scenario when a typical neutron star goes unstable through $r$-mode, particularly in the presence of differential rotation and magnetic field. 

\begin{itemize}
\item The amplitude of the $r$-mode first rises exponentially and saturates in a natural way a few hundred seconds after the mode instability sets in.

\item Small initial differential rotation causes the $r$-mode to saturate around unity. Larger and larger differential rotation suppresses the growth by substantial order of magnitude and also comparatively in shorter time scales. 

\item{Magnetic field plays a dominant role in dramatically suppressing the growth of $r$-mode amplitude. }
\item{In absence of either differential rotation or magnetic field, the angular velocity of a typical neutron star stays constant upto a time scale of few hundred seconds. Subsequently  it starts abruptly shedding in another few thousand seconds. Presence of differential rotation causes the star to shed angular velocity early. On the other hand, presence of magnetic field causes the abrupt fall of angular velocity gradually smoother.}
\item{Initial differential rotation hardly influences the critical curve for $r$-mode instability, while the magnetic field offers to \emph{shrink} the instability window available for a typical neutron star.}
\item{Presence of magnetic field tends to suppress the effects of the differential rotation parameters. Further, higher strengths of magnetic field often  \emph{erase} distinctive features induced by different values of differential rotation parameters.}
\item{The gravitational wave strain amplitude emitted due to the $r$-mode instability in a typical neutron star grows exponentially during first few hundred seconds after which it slowly falls approximately in a linar fashion.}
\item{Increased values of differential rotation cause the gravitational wave strain amplitude to peak at smaller values as well as early. Further, the magnetic field smoothens this amplitude \emph{change-over} at the peak.}
\item{The Signal-to-Noise ratio of the gravitational waves due to the models investigated in this study reveal that the Advanced LIGO detector with low values of $K$ and magnetic field upto an order of $10^{15}$ Gauss, will be able to detect $r$-mode unstable neutron stars upto a distance of 20 Megaparsecs. }

\end{itemize}

\end{document}